\documentclass[12pt,preprint]{aastex}

\shorttitle{Testing the CMB Data for Systematic Effects}
\shortauthors{Griffiths \& Lineweaver}

\begin{document}

\title{Testing the CMB Data for Systematic Effects}

\author{Louise M. Griffiths and Charles H. Lineweaver}

\affil{Department of Astrophysics and Optics, School of Physics, University of New South Wales, Sydney, NSW 2052, Australia}

\begin{abstract}
Under the assumption that the concordance $\Lambda$ cold dark matter
(CDM) model is the correct model, we test the cosmic microwave
background (CMB) anisotropy data for systematic effects by examining
the band pass temperature residuals with respect to this model.
Residuals are plotted as a function of $\ell$, galactic latitude,
frequency, calibration source, instrument type and several other
variables that may be associated with potential systematic
effects. Linear fitting to the residuals indicates no significant
identifiable systematic errors associated with these variables, except
for the case of galactic latitude.  We find evidence for a trend
associated with the absolute galactic latitude range at more than the
2-$\sigma$ level. This may be indicative of galactic contamination and
may require a 2\% reduction in the normalisation of low galactic
latitude observations.

\end{abstract}

\keywords{cosmic microwave background - cosmology: observations}

\section{INTRODUCTION}
The cosmic microwave background (CMB) power spectrum is a particularly
potent probe of cosmology.  As long as the systematic errors
associated with these observations are small, the detected signal has
direct cosmological importance. The ever-tightening network of
constraints from CMB and non-CMB observations favours a concordant
$\Lambda$ cold dark matter (CDM) model that is commonly accepted as
the standard cosmological model (Table~\ref{concord}).  Since the
anisotropy power spectrum is playing an increasingly large role in
establishing and refining this model, it is crucial to check the CMB
data for possible systematic errors in as many ways as possible.

Systematic errors and selection effects are notoriously difficult to
identify and quantify.  Calibration and/or beam uncertainties dominate
current CMB measurements and there may be lower level systematic
errors of which we are not aware \citep{page01}.  Individual
experimental groups have developed various ways to check their CMB
observations for systematic effects \citep[e.g.][]{kogut96,miller02},
including the use of multiple calibration sources, multiple frequency
channels and extensive beam calibrating observations. Internal
consistency is the primary concern of these checks.

Testing for consistency with other CMB observations is another
important way to identify possible systematic errors.  When the areas
of the sky observed overlap, this can be done by comparing CMB
temperature maps \citep[e.g.][]{ganga94a,lineweaveretal95,xu01}. When
similar angular scales are being observed one can compare power
spectra \citep[e.g.][Figure 11]{sievers02}.  A prerequisite for the
extraction of useful estimates for cosmological parameters from the
combined CMB data set is the mutual consistency of the observational
data points \citep{wangetal02a}; the best-fit must also be a good
fit. \citet{wangetal02a} and \citet{sievers02} have recently explored
the consistency of various CMB observations with respect to power
spectrum models and concluded that the CMB fluctuation data is
consistent with several minor exceptions.

Although individual observational groups vigorously test their data
sets for systematic errors, the entire CMB observational data set has
not yet been collectively tested.  Here we check for consistency of
the concordance model (Table~\ref{concord}) with respect to possible
sources of systematic error.  Under the assumption that the
concordance model is the correct model (i.e. more correct than the
best-fit to the CMB data alone), we explore residuals of the
observational data with respect to this model to see if any patterns
emerge.  We attempt to identify systematic errors in the data that may
have been ignored or only partially corrected for.

With only a few independent band power measurements the usefulness of
such a strategy is compromised by low number statistics. However, we
now have 203 measurements of band power on scales of $2<\ell<2000$
from over two dozen autonomous and semi-autonomous groups. There are
enough CMB fluctuation detections from independent observations that
subtle systematic effects could appear above the noise in regression
plots of the data residuals.  This is particularly the case when one
has a better idea of the underlying model than provided by the CMB
data alone.

The history of the estimates of the position of the CMB dipole
illustrates the idea.  Once a relatively precise direction of the
dipole was established, the positional scatter elongated in the
direction of the galactic centre could be distinguished unambiguously
from statistical scatter and more reliable corrections for galactic
contamination could be made \citep[Figure 2]{lineweaver97}.  We aim to
ascertain whether the use of the concordance model as a prior can help
to separate statistical and systematic errors in the CMB anisotropy
data.  In \S2 we discuss constraints on cosmological parameters, the
current concordance model and how simultaneously analysing
combinations of independent observational data sets can tighten
cosmological constraints. Our analytical methodology is detailed in
\S3.  In \S4 possible sources of systematic uncertainty are discussed.
In \S5 and \S6 our results are discussed and summarised.

\section{THE CONCORDANCE COSMOLOGY}
\subsection{Observational concordance} 
The CMB has the potential to simultaneously constrain a number of
cosmological parameters that are the ingredients of the hot big bang
model.  Unfortunately, particular parameter combinations can produce
indistinguishable $C_{\ell}$ spectra \citep{efbond99}.  For example,
cosmological models with different matter content but the same
geometry can have nearly identical anisotropies.  Such model
degeneracies limit parameter extraction from the CMB alone.

A number of recent analyses combine information from a range of
independent observational data sets
\citep[e.g.][]{efstetal02,lewisbridle02,sievers02,wangetal02a,wangetal02b},
enabling certain degeneracies of the individual data sets to be
resolved.  As the observational data become more precise and diverse
they form an increasingly tight network of parameter constraints.
Analyses of a variety of astrophysical observations are beginning to
refine an observationally concordant cosmological model.

\citet{efstetal02} perform a combined likelihood analysis of the power
spectra of the 2-degree Field Galaxy Redshift Survey (2dFGRS) and the
CMB anisotropies under the assumptions that the galaxy power spectrum
on large scales is directly proportional to the linear matter power
spectrum and that the initial fluctuations were adiabatic, Gaussian
and well described by power-laws with scalar and tensor indeces of
$n_s$ and $n_t$. 11 cosmological parameter combinations are
simultaneously considered; the curvature parameter $\Omega_{\kappa}$,
the contribution to the overall energy density from the cosmological
constant $\Omega_{\Lambda}$, $\omega_b = \Omega_bh^2$ (where
$\Omega_b$ is the baryon energy density and $h$ the Hubble parameter),
$\omega_c = \Omega_ch^2$ (where $\Omega_c$ is the energy density of
CDM), the amplitude of scalar perturbations $A_s$, the ratio of tensor
to scalar perturbations $A_t/A_s$, the spectral index of scalar
perturbations $n_s$, that of tensor perturbations $n_t$, the
reionisation optical depth $\tau$, the shape parameter that defines
the turn-over of the matter power spectrum $\Gamma \simeq \Omega_mh$
(where $\Omega_m$ is the energy density of non-relativistic matter)
and the bias parameter $b$. The big bang nucleosynthesis (BBN)
constraint on the baryon content \citep{burlesetal01} is added to the
analysis as a prior. The results for the CMB alone and combined
analyses are given in Table~\ref{concord}.

\citet{wangetal02a} perform a similar analysis using the CMB data, the
decorrelated linear power spectrum extracted from the PSC$z$ survey
\citep{hamteg02} and the Hubble Key Project (HKP) prior for $h$
\citep{freedmanetal01}.  Instead of assuming that the dark matter
contribution to the energy density is entirely composed of cold dark
matter, they explore the parameter combination $\omega_d =
\Omega_dh^2$ and introduce a new parameter, $f_{\nu} =
\Omega_{\nu}/\Omega_d$, that is the fraction of dark matter that is
hot.  Their constraints from the CMB alone and combined analyses are
also given in Table~\ref{concord}.

Another recent analysis of the CMB data set is that of
\citet{sievers02}.  They perform a likelihood analysis over 7
cosmological parameters ($\Omega_{\rm total} = 1 - \Omega_{\kappa}$,
$\Omega_{\Lambda}$, $\omega_b$, $\omega_c$, $A_s$, $n_s$, $\tau$)
applying a sequence of increasingly strong prior probabilities
successively to the likelihood functions.  These are a flat prior
$\Omega_{\kappa} = 0$ in accordance with the predictions of the
simplest inflationary scenarios, a large scale structure prior that
involves a constraint on the amplitude $\sigma_8^2$ and shape of the
matter power spectrum, the HKP prior for $h$ and the
$\Omega_m-\Omega_{\Lambda}$ priors from supernova type Ia (SNIa)
observations \citep{riess98,perlmutter99}.  Their results for the CMB
alone and the CMB+priors likelihood analyses are also given in
Table~\ref{concord}.

\citet{lewisbridle02} implement a Markov chain Monte Carlo method to
constrain 9 parameter flat models using a subsample of the CMB data
set together with the BBN, HKP and SNIa priors.  They also perform a
combined analysis with the 2dFGRS.  The most recent joint analysis of
the CMB data set and the 2dFGRS is that of \citet{wangetal02b}.  The
results of both these analysis are also given in Table~\ref{concord}.

Due to the degeneracies in the anisotropy power spectrum, the CMB
alone is only able to provide weak constraints on particular
cosmological parameters. Combining the CMB constraints with the
results of independent observational data sets can tighten these
constraints.  The results of the joint likelihood analyses discussed
in this section suggest the observationally concordant cosmology;
$\Omega_{\kappa}\simeq0$, $\Omega_{\Lambda}\simeq 0.7$ ($\Omega_{m} =
\Omega_b+\Omega_c\simeq 0.3$), $\Omega_b h^2 \simeq0.02$, $n_s \simeq
1$ and $h \simeq 0.68$ with $A_t$, $\tau$ and $\Omega_{\nu}$ taken to
be zero.  With more precise and diverse cosmological observations, the
ability of the standard $\Lambda$CDM cosmology to describe the
observational universe will be extended and tested for
inconsistencies.

\subsection{Goodness of fit of the concordance cosmology to the CMB}
We perform a simple $\chi^{2}$ calculation (see Appendix~A) to
determine the goodness-of-fit of this new standard $\Lambda$CDM
cosmology to the CMB, employing the band power temperature
measurements in Table~\ref{obsdat} and their associated window
functions.  We limit our analysis to $2<\ell<2000$ because secondary
anisotropy contributions, such as the Sunyaev-Zel'dovich
\citep{sunzel70} effect, may dominate at $\ell>2000$
\citep[e.g.][]{bondetal02}.  The model radiation angular power spectrum is
calculated using {\sc cmbfast} \citep{selzal96}.  However, rather than
adopting the {\sc cmbfast} COBE-DMR normalisation, we implement the
numerical approximation to marginalisation (see Appendix A) to find
the optimal normalisation of the theoretical model to the full
observational data set.  We also similarly treat the beam
uncertainties of BOOMERanG98 and MAXIMA1 as given by
\citet{lesgorgues01} and the calibration uncertainties associated with
the observations, treating them as free parameters with Gaussian
distributions about their nominal values (see Eq.~\ref{e:chisq}).

The minimised $\chi^2$ for the concordance model is 174.2. In order to
determine how good a fit this model is to the observational data we
need to know the number of degrees of freedom of the
analysis. Although 203 degrees of freedom are provided by the number
of observational data points (assuming they are uncorrelated), these
are reduced by the number of concordance parameters that are
constrained using the CMB data alone.  The flatness of the concordance
model ($\Omega_{\kappa} \simeq 0$) and the scale invariance of the
primordial power spectrum of scalar perturbations ($n_s \simeq 1$) are
extracted almost entirely from the CMB data.  The remaining concordant
parameters are more strongly constrained by non-CMB observations.  We
therefore estimate that 2 degrees of freedom should be subtracted from
the original 203.

Within our analysis we marginalise over a number of nuisance
parameters.  We fit for 23 individual calibration constants, 2 beam
uncertainties (those of BOOMERanG-98 and MAXIMA-1) and an overall
normalisation.  Thus a further 26 degrees of freedom must be
subtracted leaving 175 degrees of freedom.  The $\chi^2$ per degree of
freedom is then 1.0, indicating that the concordance cosmology
provides a good fit to the CMB data alone.

Data correlations other than the correlated beam and calibration
uncertainties of individual experiments, that we take to have no
inter-experiment dependence, are not considered in our analysis.
Including such correlations would further reduce the number of degrees
of freedom, increasing the $\chi^2$ per degree of freedom.  However,
our result is in agreement with the joint likelihood analyses that
find that the cosmological model that best fits the CMB data is a
better fit at the 1 or 2 $\sigma$ level than fit to the concordance
model \citep{wangetal02a}.

We find the normalisation of the concordance model to the full CMB
data set to be $Q_{10} = 16.3 \pm 0.1 \,\mu$K, where $Q_{10}$ is defined
through the relation \citep{linbar98},
\begin{equation}
10(10+1)C_{10} = \frac{24\pi}{5} \frac{Q_{10}^2}{T_{CMB}^2} \,.
\end{equation}
The normalised concordance model is plotted with the calibrated and
beam corrected observational data in Figure~\ref{clplotcol}. It is
difficult to distinguish the most important measurements because there
are so many CMB data points on the plot and it is dominated by those
with the largest error bars.  Therefore, for clarity, we bin the data
as described in Appendix~B.  The binned observations are plotted with
the concordance cosmology in Figure~\ref{bin_clplotcol}.  The linear
$x$-axis emphasises the detail at small angular scales, clarifying
measurements of the acoustic peaks.

\section{EXAMINING THE RESIDUALS}
Our analysis is based on the assumption that the combined cosmological
observations used to determine the concordance model are giving us a
more accurate estimate of cosmological parameters, and therefore of
the true $C_{\ell}$ spectrum, than is given by the CMB data
alone. Under this assumption, the residuals of the individual observed
CMB band powers and the concordance $\Lambda$CDM model become tools to
identify a variety of systematic errors.  To this end, we create
residuals, $R_i$, of the observed band power temperature anisotropies
$\delta T_i^{\rm obs}\pm\sigma^{\rm obs}_i$ with respect to the
concordant band powers $\delta T_i^{\rm th}$ such that,
\begin{equation}
\label{e:resid}
R_i = \frac{\delta T_i^{\rm obs}-\delta T_i^{\rm th}}{\delta T_i^{\rm
th}} \pm \frac{\sigma^{\rm obs}_i}{\delta T_i^{\rm th}} \,.
\end{equation}

Systematic errors are part of the CMB band power estimates at some
level. We examine our data residuals as functions of the instrument
type, receivers, scan strategy and attitude control.  Possible sources
of systematic uncertainty are discussed in the following section and
the instrumental and observational details that may be associated with
systematic errors are listed in Table~\ref{obstech}. We look for any
linear trends that may identify systematic effects that are correlated
with these details of the experimental design.  We quote the $\chi^2$
per degree of freedom of the best fitting line and the significance of
the fit for each regression in Table~\ref{resres}. If the analysis
determines that a linear trend can produce a significantly improved
fit in comparison to that of a zero gradient line (zero-line) through
the data, it may be indicative of an unidentified systematic source of
uncertainty.

The zero-line through all the residual data gives a $\chi^2$ of 174.2.
The analysis that determines the goodness-of-fit of the concordance
model to the CMB data has 175 degrees of freedom. If the gradient and
intercept were independent of the parameters varied to produce the
concordance fit, the degrees of freedom would be further reduced by 2.
However, the intercept of any line that fits the residual data will
depend on the normalisation of the concordance model. We therefore
subtract only one further degree of freedom, giving 174 degrees of
freedom.

The best fitting zero-line fit to all the residual data has a $\chi^2$
per degree of freedom of 1.0 ($=174.2/174$). In order to determine the
significance of a better fit provided by a linear trend, an
understanding of the statistical effects of introducing the 2
parameters to the line-fitting analysis is required.  For a
2-dimensional Gaussian distribution, the difference between the
$\chi^2$ of the best-fit model and a model within the 68\% confidence
region of the best-fit model is less than 2.3 and for a model that is
within the 95\% confidence region of the best-fit model, this
difference is less than 6.17 \citep{numrec}.  Our 68\% and 95\%
contours in Figures~\ref{lresid} to \ref{pdresid} are so defined.  The
further the horizontal concordance zero-line is from the best fitting
slope, the stronger the indication of a possible systematic error.

\section{POSSIBLE SOURCES OF SYSTEMATIC UNCERTAINTY}
\subsection{Foregrounds}
If foreground emission is present, it will raise the observed power.
Galactic and extra-galactic signals from synchrotron, bremsstrahlung
and dust emission have frequency dependencies that are different from
that of the CMB \citep[e.g.][]{tegef96}. If such contamination is
present in the data, it may be revealed by a frequency dependence of
the residuals (Figure \ref{nuresid}).  Multiple frequency observations
provide various frequency lever-arms that allow individual groups to
identify and correct for frequency dependent contamination.
Experiments with broad frequency coverage may be better able to remove
this contamination than those with narrow frequency coverage.  We
therefore examine the residuals as a function of the frequency
lever-arm $(\nu_{\rm max}-\nu_{\rm min})/\nu_{\rm main}$ (Figure
\ref{deltanuresid}).

Observations taken at lower absolute galactic latitudes, $|b|$, will
be more prone to galactic contamination. In Figures \ref{gallatresid},
\ref{gallongresid} and \ref{gallatresidb} we check for this effect by
examining the residuals as a function of $|b|$
(Figures~\ref{gallatresid} and \ref{gallatresidb}) and galactic
longitude (Figure~\ref{gallongresid}).  Less likely would be a signal
associated with the narrowness of the band pass of the main frequency
channel $\Delta\nu_{\rm main}/\nu_{\rm main}$ (Figure \ref{dnuresid}).

\subsection{Angular Scale-dependent effects}
We examine scale-dependent uncertainties by plotting the residuals as
a function of $\ell$ (Figure \ref{lresid}).  The shape of the window
function is most critical when the curvature of the power spectrum is
large (at the extrema of the acoustic oscillations). We therefore
explore the residuals as a function of the narrowness of the filter
functions in $\ell$ space $\Delta \ell/\ell$ (Figure \ref{dlresid}).

The area of the sky observed determines the lowest $\ell$ probed while
the beam size $\theta_{\rm beam}$ determines the highest $\ell$.  The
resolution of the instrument and the pointing uncertainty become
increasingly important as fluctuations are measured at smaller angular
scales.  Small beams may be subject to unidentified smearing effects
that may show up as a trend in the residual data with respect to
$\theta_{beam}/ \ell_{eff}$.  Thus we examine the residuals as a
function of the area of sky probed (Figure \ref{arearesid}),
$\theta_{beam}/ \ell_{eff}$ (Figure \ref{resresid}) and pointing
uncertainty (Figure \ref{pointresid}) to look for hints of systematic
errors associated with these factors.

\subsection{Calibration}
To analyse various experiments, knowledge of the calibration
uncertainty of the measurements is necessary.  Independent
observations that calibrate off the same source will have calibration
uncertainties that are correlated at some level and therefore a
fraction of their freedom to shift upwards or downwards will be
shared.  For example, ACME-MAX, BOOMERanG97, CBI, MSAM, OVRO, TOCO and
CBI all calibrate off Jupiter, so part of the quoted calibration
uncertainties from these experiments will come from the brightness
uncertainty of this source.  The remainder will be due to detector
noise and sample variance and should not have any such
inter-experiment correlations. \citet{wangetal02a} perform a joint
analysis of the CMB data making the approximation that the entire
contribution to the calibration uncertainty from Jupiter's brightness
uncertainty is shared by the experiments that use this calibration
source.  The true correlation will be lower since the independent
experiments observed Jupiter at different frequencies.

Inter-experiment correlations are not considered in our analysis,
since we are unable to separate out the fraction of uncertainty that
is shared by experiments.  Instead we test for any calibration
dependent systematics by examining the data residuals with respect to
the calibration source (Figure \ref{csresid}). We note that including
correlations between data points would reduce the number of degrees of
freedom of our $\chi^2$ analysis.

\subsection{Instrument type, platform and altitude}
The experiments use combinations of 3 types of detector that operate
over different frequency ranges.  We classify the data with respect to
their instrument type; HEMT interferometers (HEMT/Int), HEMT amplifier
based non-interferometric instruments (HEMT), HEMT based amplifier and
SIS based mixer combination instruments (HEMT/SIS), bolometric
instruments and bolometric interferometers (Bol/Int). We check for
receiver specific systematic effects by plotting the residuals as a
function of instrument type (Figure \ref{typeresid}).

Water vapour in the atmosphere is a large source of contamination for
ground based instruments.  There may also be systematic errors
associated with the temperature and stability of the thermal
environment.  We therefore explore instrument altitude (Figure
\ref{altresid}) and platform (Figure \ref{platresid}) dependencies of
the data residuals.

\subsection{Random controls}
We use a number of control regressions to check that our analysis is
working as expected.  To this end, the residuals are examined with
respect to the publication date of the band power data (Figure
\ref{pdresid}), the number of letters in the first author's surname
(Figure \ref{nletresid}) and the affiliation of the last author
(Figure \ref{affilresid}).  We expect the line fitted to these control
regressions to be consistent with a zero-line through the residual
data.  Any significant improvement provided by a linear fit to these
residuals may be indicative of a problem in the software or
methodology.

\section{RESULTS}
For the regressions plotted, the residual data is binned as described
in Appendix~B so that any trends can be more effectively visualised.
Since the data binning process may wash out any discrepancies between
experiments, the linear fit analyses are performed on the unbinned
data residuals.  In Figures~\ref{lresid} to \ref{pdresid}, the line
that best-fits the data is plotted (solid white) and the 68\% (dark
grey) and 95\% (light grey) confidence regions of the best-fit line
are shaded.  For each plot, we report the $\chi^2$ and the $\chi^2$
per degree of freedom for the best-fit line, the probability of
finding a model that better fits the data and comment on the
significance of the deviation of the zero-line (dashed black).

Our results are listed in Table~\ref{resres} and imply that the most
significant linear trend observable in the residuals is with respect
to the absolute galactic latitude $|b|$ of the observations (see
Figure~\ref{gallatresid}).  This trend is not eliminated by the
removal of any one experiment and may be indicative of a source of
galactic emission that has not been appropriately treated. The
weighted average of points $|b| > 40^{\circ}$ is $\sim - 1\%$, while
it is $\sim + 1\%$ for $|b| < 40^{\circ}$.  If this is due to galactic
contamination, then the normalisation Q$_{10}$ may have to be reduced
to 16.1 $\mu$K.

For this regression, the errors in both the $y$ and $x$ direction are
used in the fit. We have defined $|b|$ to be that of the centre of the
observations and the uncertainties to extend to edges of the range.
This allows the observations some freedom of the $x$-coordinate in the
line-fitting analysis and may over-weight those detections that span
small ranges in absolute galactic latitude. It is therefore also
interesting to examine the residuals with respect to the central $|b|$
to determine the significance of the trend with the $x$-coordinate
freedom removed (see Figure~\ref{gallatresidb}). The most plausible
galactic latitude regression will be somewhere between the two plots.

Removing the $x$-coordinate freedom removes the significance of the
trend. This result implies that experiments that observe over small
ranges in galactic latitude are dominating the trend and we therefore
can not simply correct for the systematic that is implied in
(Figure~\ref{gallatresid}).  The comparison of rms levels in galactic
dust \citep{finkbeiner99} and
synchrotron\footnote{http://astro.berkeley.edu/dust} maps over the
areas of CMB observations may help to clarify the interpretation of
the trend.  Such a technique has recently been applied to the MAXIMA1
data \citep{jaffeetal03} but has yet to be performed on the full CMB
data set.

Other plots also show some evidence for systematic
errors. Figures~\ref{bin_clplotcol} and \ref{lresid} indicate that the
6 bins between $1000\le\ell\le 2000$ prefer a lower normalisation.
This could be due to underestimates of beam sizes or pointing
uncertainties or unidentified beam smearing effects at high $\ell$ for
small beams.  However, Figures \ref{resresid} and \ref{pointresid}
show no evidence for any trends, although limiting the pointing
uncertainty analysis to the 5 points with the largest uncertainties
would indicate a trend, suggesting that the largest pointing
uncertainties may have been underestimated.

\section{SUMMARY}
Although individual observational groups vigorously test their data
sets for systematic errors, the entire CMB observational data set has
not yet been collectively tested.  Under the assumption that the
concordance model is the correct model, we have explored residuals of
the observational data with respect to this model to see if any
patterns emerge that may indicate a source of systematic error.

We have performed linear fits on the residual data with respect to
many aspects of the observational techniques and, for the majority, we
have found little or no evidence for any trends.  However, there is
significant evidence for an effect associated with $|b|$ that is not
eliminated with the exclusion of any one data set. The data prefers a
linear trend that is inconsistent at more than 95\% confidence with a
zero gradient line through the residuals. The best-fit line to the
data suggests that CMB observations made closer to the galactic plane
may be over-estimated by approximately 2\%.  A more detailed analysis
of galactic dust and synchrotron maps may clarify the source of the
indicated systematic uncertainty.

\acknowledgments LMG thanks Martin Kunz for useful discussions and is
grateful to the University of Sussex where part of the work was
carried out.  LMG acknowledges support from the Royal Society and
PPARC. CHL acknowledges a research fellowship from the Australian
Research Council.

\clearpage

\appendix

\section{$\chi^2$ MINIMISATION METHODOLOGY}
For the observational power spectrum data quoted in the literature
(Table~\ref{obsdat}), individual $C_{\ell \,}$s are not estimated,
rather band powers are given that average the power spectrum through a
filter, or {\em window function}. Each theoretical model must
therefore be re-expressed in the same form before a statistical
comparison can be made.

The theoretical model can be re-expressed as $\delta T_{\ell_{\rm
eff}}^{\rm th}$, using the method of \citet{linetal97}. Boltzmann codes
such as {\sc cmbfast} \citep{selzal96} output theoretical power
spectra in the form,
\begin{equation}
d_1({\ell})=\frac{\ell(\ell \, + \, 1)}{2\pi} \, C_{\ell}^{\rm theory}
\times {\rm normalisation} \,.
\end{equation}
Since the $C_\ell \,$s are adimensional, they are multiplied by $\bar
T_{\rm CMB}^2 \simeq (2.725 \, {\rm K})^2$ \citep{mather99} to
express them in Kelvin,
\begin{equation}
d_2({\ell}) \, = \, \bar T_{\rm CMB}^2 \, d_1({\ell}) \,. 
\end{equation}

The sensitivity of each observation (denoted $i$) to a particular
$\ell$ is incorporated using the observational window function
$W_{\ell}$,
\begin{equation}
d_3(i, \, \ell) \, = \, d_2({\ell}) \times \frac{(2 \,
\ell \, +1 \, ) \, W_{\ell}^{i}}{2\, \, \ell \, (\ell \, + \, 1)} \,.
\end{equation}
The contribution from the model to the $i^{\rm th}$ observational
band-power is determined and the influence of the window function
removed,
\begin{equation}
D^{\rm th}_i \, = \, \frac{\sum_{\ell = 2}^{\ell_{\rm max}} d_3(i, \, \ell)}{I(i)} \,,
\end{equation}
where $I$($i$) is the logarithmic integral of the window function, 
\begin{equation}
I(i) \, = \sum_{\ell=2}^{\ell_{\rm max}}\frac{(2 \,
\ell \, +1 \, ) \, W_{\ell}^{i}}{2\, \, \ell \, (\ell \, + \, 1)}
\,.
\end{equation}
$\delta T_{\ell_{\rm eff}}^{\rm th}(i)$ is then,
\begin{equation}
\delta T_{\ell_{\rm eff}}^{\rm th}(i) = \sqrt{D^{\rm th}_i} \,,
\end{equation}
which can be statistically compared with the $i^{\rm th}$
observational band-power measurement $\delta T_{\ell_{\rm eff}}^{\rm
obs}(i) \equiv \sqrt{D^{\rm obs}_i}$ given in Table~\ref{obsdat}.

The assumption that the CMB signal is a Gaussian random variable
enables analysis via a likelihood procedure. Due to the non-Gaussian
distribution of the uncertainty in the band-power measurements, an
accurate calculation of the likelihood function $L$ is
non-trivial. However, approximations to the true likelihood have been
derived \citep{bondetal00,bartlett00}.  For example, the
\citet{bondetal00} offset lognormal formalism is implemented in the
publicly available {\sc radpack} package.  Unfortunately, the
information necessary to implement this formalism has not yet been
published by all observational groups.  Therefore, in order to
statistically analyse the complete CMB observational data set, we make
the assumption that $L$ is Gaussian in $D_i$.  Then,
\begin{equation}
\label{e:chisq}
\chi^2 \equiv -2\ln{L} = \sum_i{\left(\frac{D^{\rm th}_i - D^{\rm obs}_i}{\sigma^{\rm obs}_i}\right)^2}.
\end{equation}

The normalisation of the primordial power spectrum is not predicted by
inflationary scenarios and therefore the normalisation of the
concordance model to the full CMB observational data set is a free
parameter.  Unless we are particularly interested in the amplitude of
primordial fluctuations, we can treat the model normalisation $A$ as a
nuisance parameter.  Assuming a Gaussian likelihood, marginalisation
can be approximated numerically for the power spectrum normalisation
by computing the $\chi^2$ statistic of the concordance model for a
number of discrete steps over the normalisation range.  The
normalisation that minimises the $\chi^2$ can thereby be determined
for a particular theoretical model.

The CMB measurements have associated calibration uncertainties that
allow data from the same instrument that is calibrated using the same
source to shift collectively upwards or downwards. The calibration
uncertainties $\sigma_c$ are given in Table~\ref{obsdat}.  The
observational band-powers are multiplied by a calibration factor $C$
that can be treated as a nuisance parameter with a Gaussian
distribution about 1.  This introduces an additional $\chi^2$ term to
Eq.~\ref{e:chisq} for each experiment that has an associated
calibration uncertainty (see Eq.~\ref{e:chisq1}).

Additionally, the BOOMERanG98 and MAXIMA1 data sets have quantified
beam and pointing uncertainties.  The combined beam plus pointing
uncertainty for each experiment introduces an additional term to
Eq.~\ref{e:chisq} that is a function of $B$. $B$ can be treated as a
nuisance parameter, with a Gaussian distribution in $B\sigma_{b, \,
i}$ about 0 (see Eq.~\ref{e:chisq1}). \citet{lesgorgues01} give
fitting functions for the combined beam plus pointing uncertainty in
$D^{\rm obs}_i$ for each of these experiments; $\sigma_{b, \, \ell} =
0.43 \times 10^{-6}\ell^2$ for BOOMERanG98 and $\sigma_{b, \, \ell} =
10^{-6}\ell^{1.7}$ for MAXIMA1.

The nuisance parameters are incorporated into Eq.~\ref{e:chisq} to give,
\begin{equation}
\label{e:chisq1}
\chi^2 = \sum_i{\left(\frac{A \, D^{\rm th}_i -
\left(C_k+B_k\sigma_{b,
\,i}\right)D^{\rm obs}_i}{\sigma_i}\right)^2} +
\sum_k \left(\frac{C_k - 1}{\sigma_{c, \,k}}\right)^2 + \sum_k
\left(B_k\right)^2 \,,
\end{equation}
where the sum on $k$ is over the number of independent observational
data sets.  The numerical marginalisation approximation that takes
discrete steps over nuisance parameters to find those that minimise
the $\chi^2$ statistic is a computationally intensive technique.  Each
nuisance parameter introduced to the analysis increases the number of
iterations of the $\chi^2$ minimisation process by a factor of the
number of discrete steps taken over the nuisance parameter range. The
assumption that the uncertainties in $D^{\rm obs}$ are Gaussian
distributed enables an analytical approximation to be made that is
computationally less time-consuming.  However, since we are only
interested in one cosmological model, the concordance $\Lambda$CDM
model, we do not implement an analytical approximation here.

\section{OBSERVATIONAL DATA BINNING}
\begin{figure}
\plotone{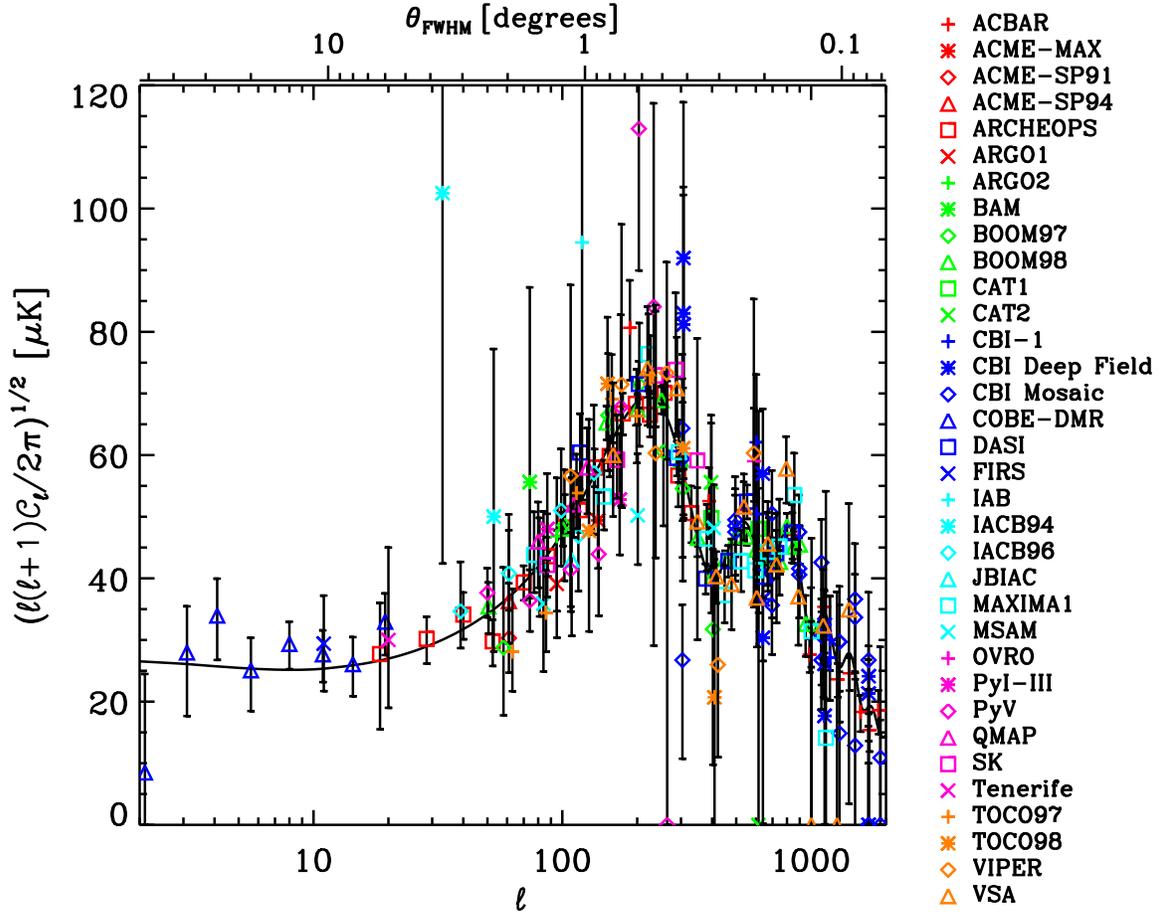}
\caption{\label{clplotcol} The concordance cosmology
(Table~\ref{concord}) normalised to the full CMB data set is plotted
with the recalibrated and, for BOOMERanG98 and MAXIMA1, beam corrected
CMB observational data given in Table~\ref{obsdat} that spans the
scales $2<\ell<2000$.}
\end{figure}

\begin{figure}
\plotone{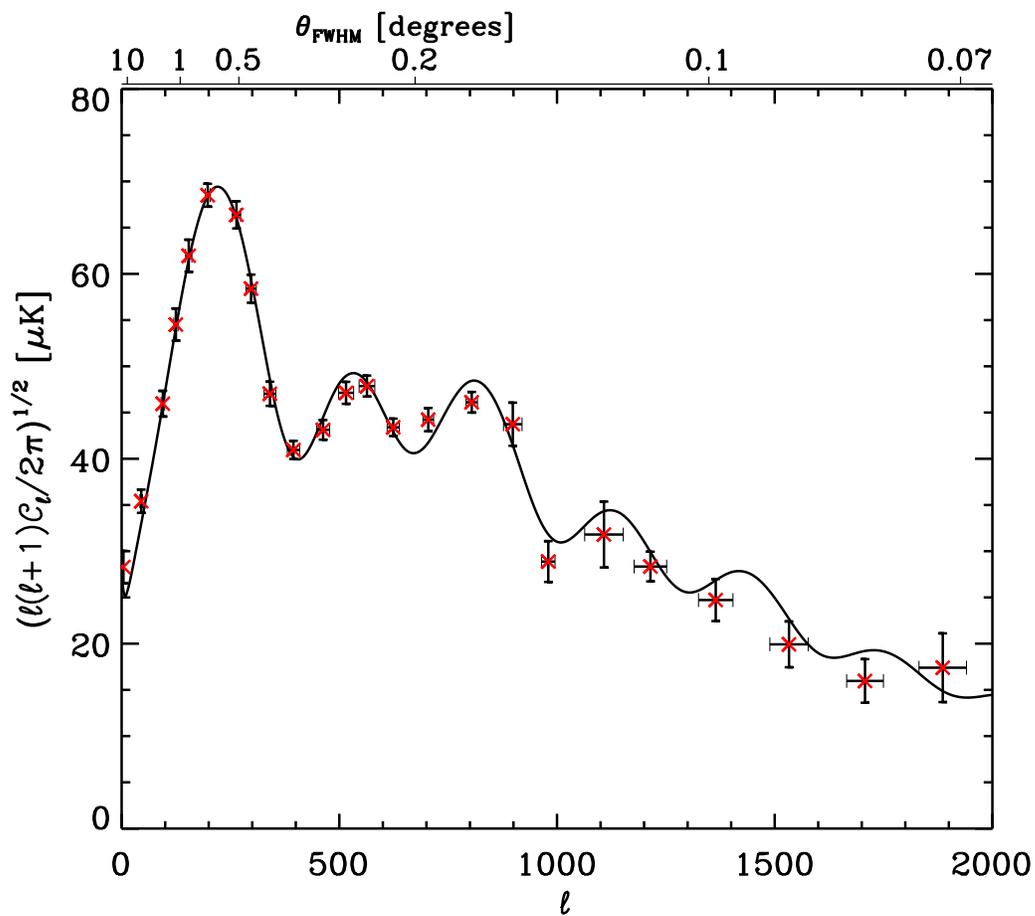}
\caption{\label{bin_clplotcol} The concordance cosmology
(Table~\ref{concord}) normalised to the full CMB data set is plotted
with the binned observational data.  The binning methodology is given
in Appendix B.  All statistical analyses detailed in this paper are
performed on the raw, unbinned data given in Table~\ref{obsdat}.}
\end{figure}

\begin{figure}
\plotone{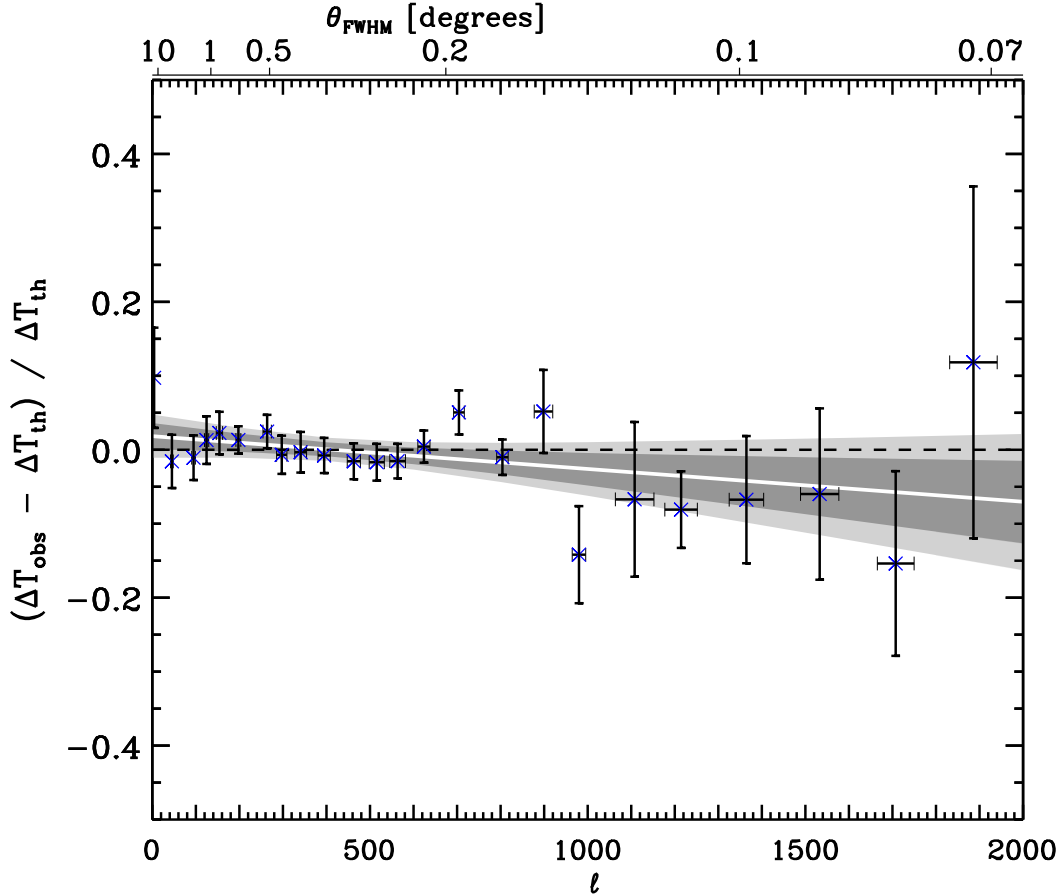}
\caption{\label{lresid}CMB data residuals plotted against $\ell$
(bottom $x$-axis) and angular scale (top $x$-axis). The fitting
routine uses the $\ell$-range as an uncertainty in the $\ell$
value. The $\chi^2$ for the fit of the line to the data is 170.6 with
a $\chi^2$ per degree of freedom of 0.98 for the 174 degrees of
freedom. The probability of finding a line that better fits the data
is 44\% so the best-fit line is a better fit to the data than the
zero-line.  The zero-line is slightly outside the border of the 68\%
confidence region for the best-fit line so there is evidence for a
trend in this regression plot at slightly greater than the 1-$\sigma$
level.}
\end{figure}

\begin{figure}
\plotone{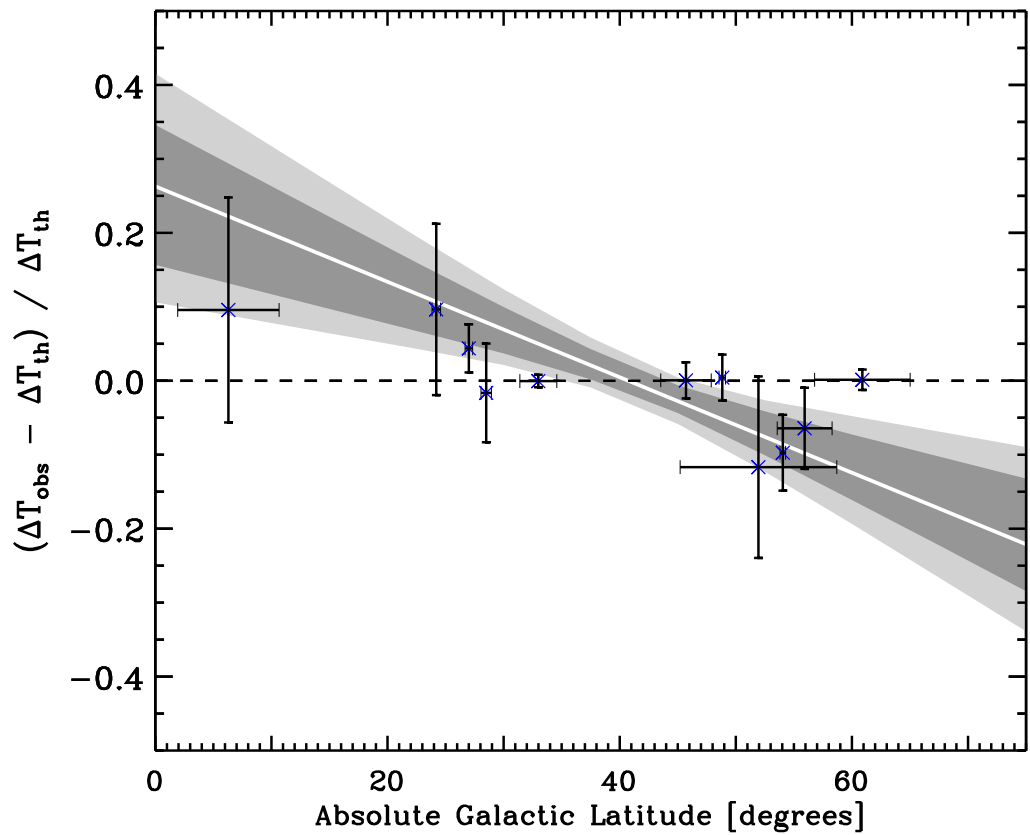} 
\caption{\label{gallatresid}CMB data residuals plotted against
absolute galactic latitude $|b|$. The fitting routine assumes that the
uncertainty in $|b|$ extends to the limits of the $|b|$ range. The
$\chi^2$ for the fit of the line to the data is 153.7 with a $\chi^2$
per degree of freedom of 0.88 for the 174 degrees of freedom. The
probability of finding a line that better fits the data is 14\% so the
best-fit line is a much better fit to the data than the zero-line.
The zero-line is outside the 95\% confidence region for the best-fit
line so there is a more than 2$\sigma$ trend in this regression plot
that may be indicative of a systematic error associated with absolute
galactic latitude.}
\end{figure}

\begin{figure}
\plotone{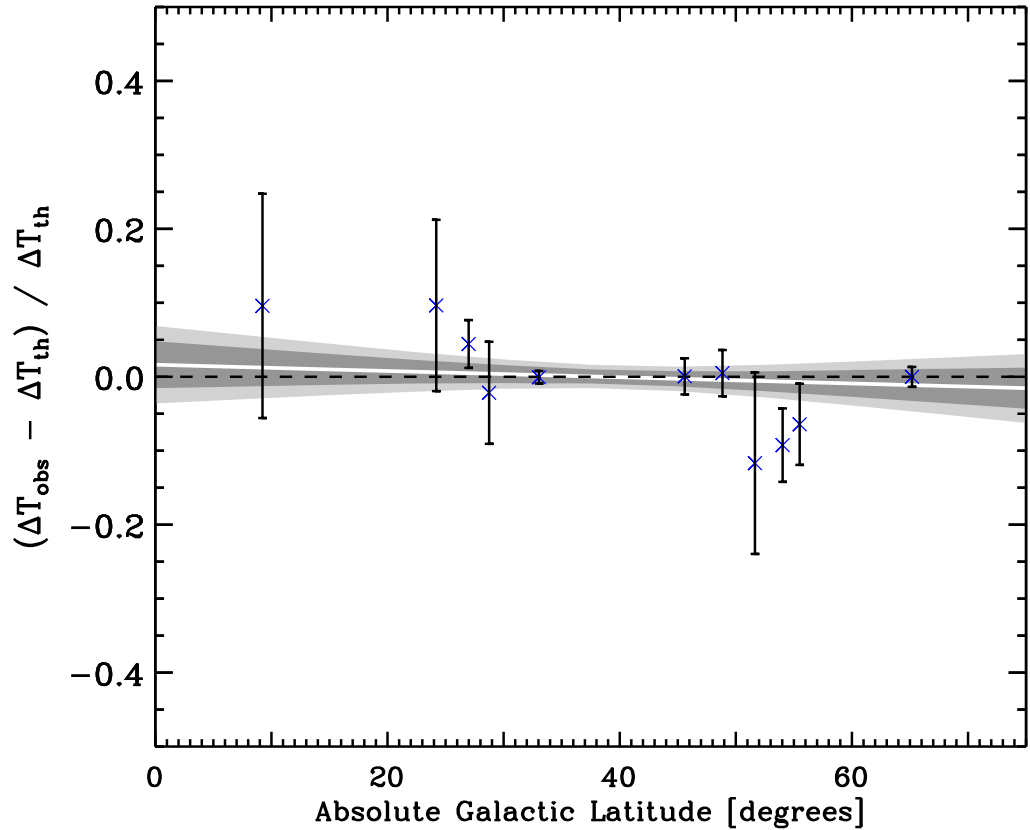}
\caption{\label{gallatresidb}CMB data residuals plotted against the
central absolute galactic latitude (i.e. the $x$-coordinate freedom
has been removed from the fit). The $\chi^2$ for the fit of the line
to the data is 173.5 with a $\chi^2$ per degree of freedom of 1.00 for
the 174 degrees of freedom. The probability of finding a line that
better fits the data is 50\% so the best-fit line improves the fit
beyond that of the zero-line only very slightly. Thus, we can conclude
that there is no evidence for a trend in this regression plot.}
\end{figure}

\begin{figure}
\plotone{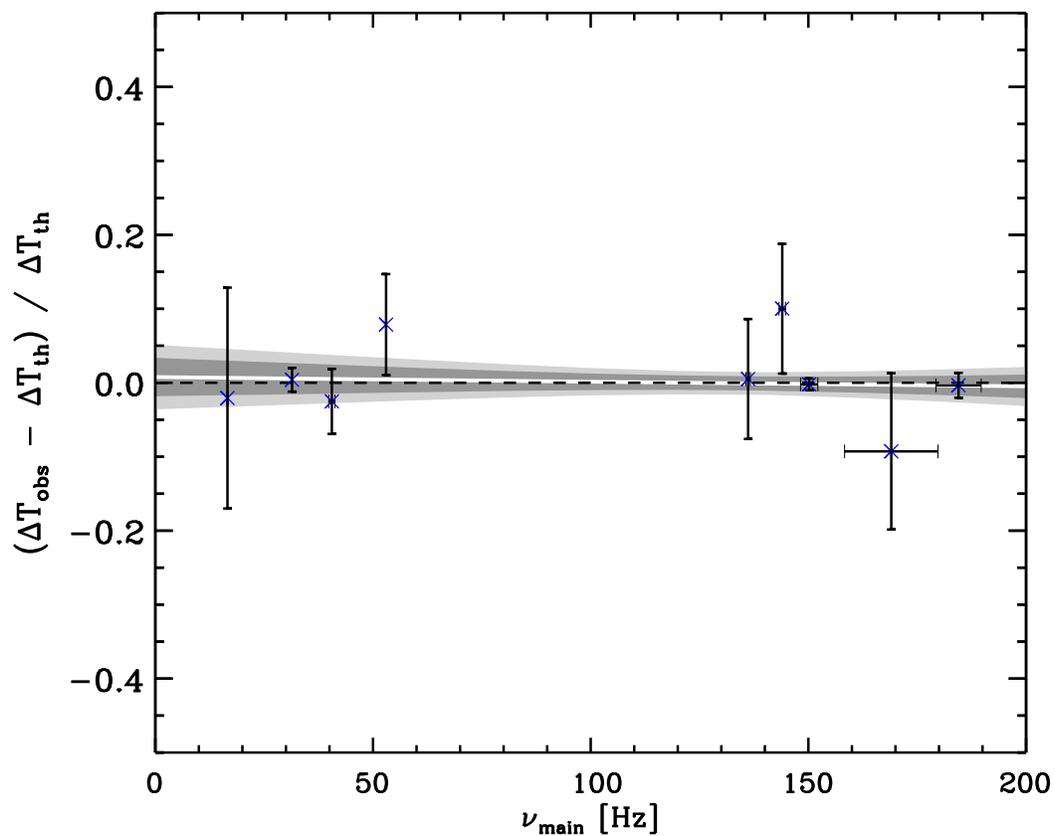} 
\caption{\label{nuresid}CMB data residuals plotted against the main
frequency of individual instruments $\nu_{\rm main}$. The $\chi^2$ for
the fit of the line to the data is $173.9$ with a $\chi^2$ per degree
of freedom of 1.00 for the 174 degrees of freedom. The probability of
finding a line that better fits the data is 51\% so the best-fit line
improves the fit beyond that of the zero-line only very
slightly. Thus, there is no evidence for a trend
in this regression plot.}
\end{figure}

\begin{figure}
\plotone{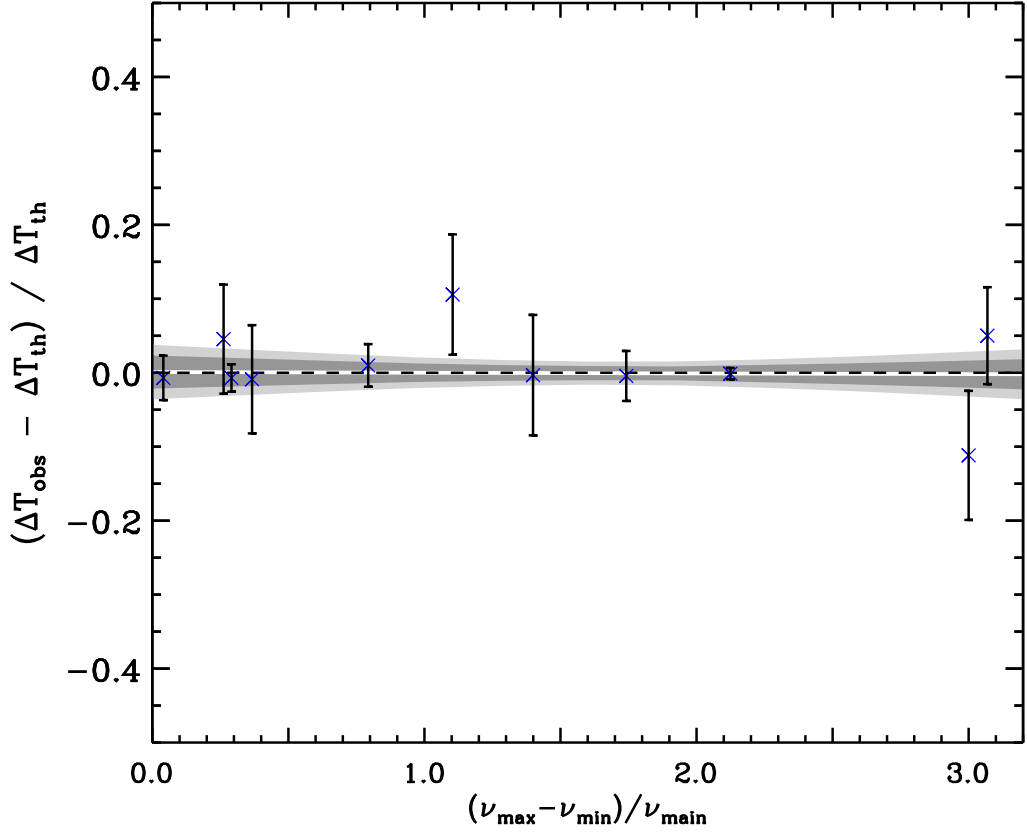}
\caption{\label{deltanuresid}CMB data residuals plotted against the
lever-arm in frequency $(\nu_{\rm max}-\nu_{\rm min})/\nu_{\rm
main}$. The $\chi^2$ for the fit of the line to the data is $174.2$
with a $\chi^2$ per degree of freedom of 1.00 for the 174 degrees of
freedom. The probability of finding a line that better fits the data
is 52\% so the best-fit line does not improve the fit beyond that of
the zero-line.  Thus, there is no evidence for a
trend in this regression plot.}
\end{figure}

\begin{figure}
\plotone{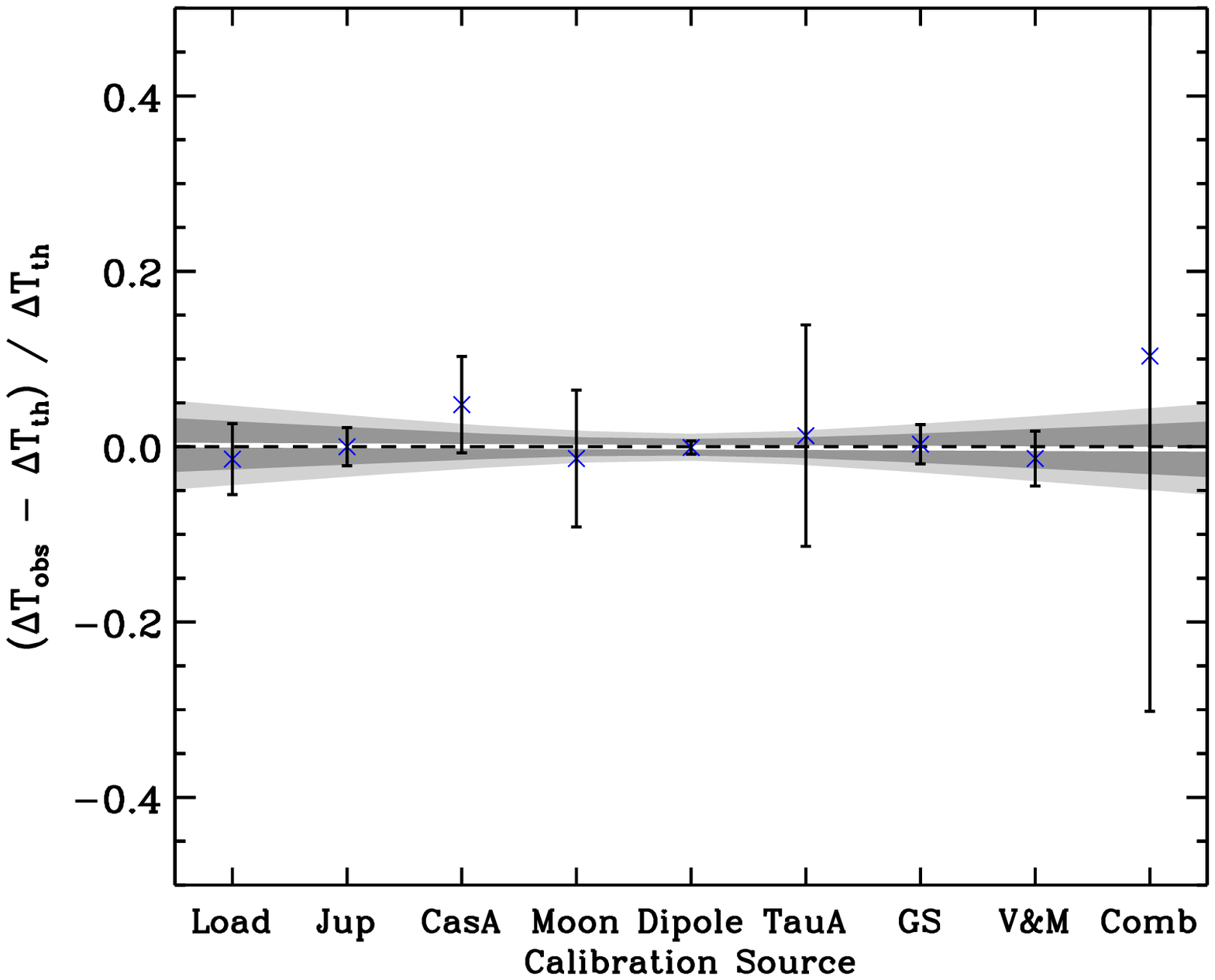} 
\caption{\label{csresid}CMB data residuals plotted against calibration
source. The $\chi^2$ for the fit of this line to the data is $174.2$
with a $\chi^2$ per degree of freedom of 1.00 for 174 degrees of
freedom.  The probability of finding a line that better fits the data
is 52\% so the best-fit line does not improve the fit beyond that of
the zero-line.  Thus, there is no evidence for a
trend in this regression plot.}
\end{figure}

\begin{figure}
\plotone{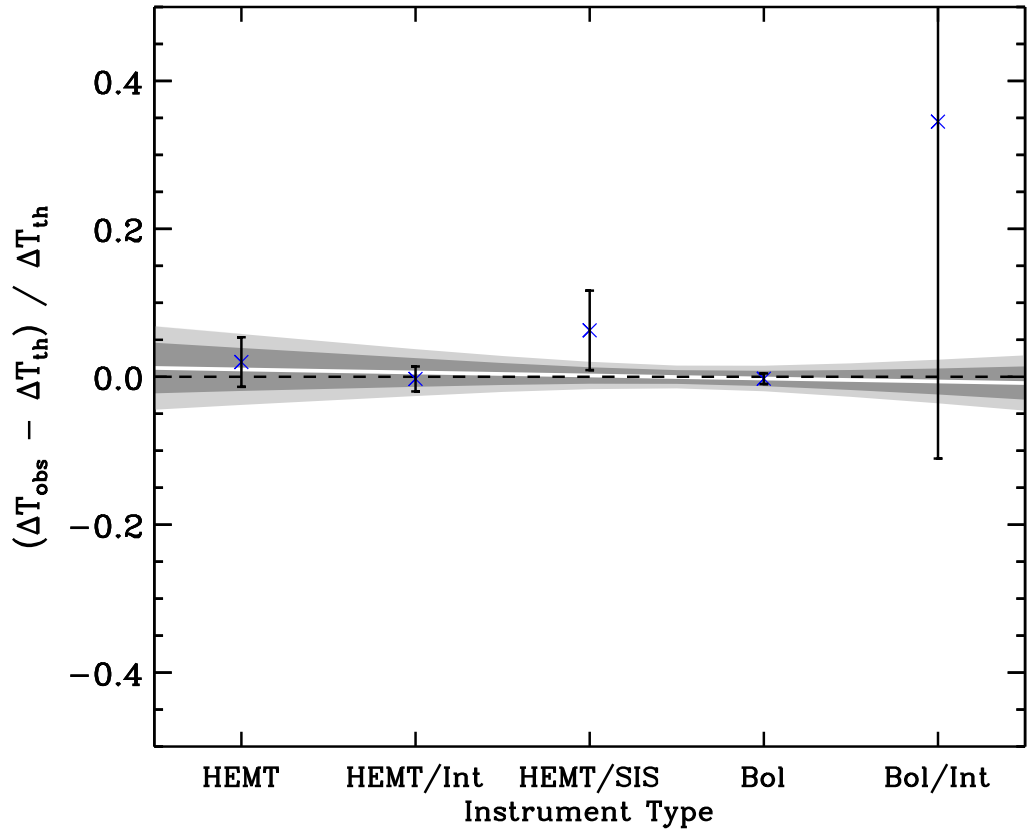} 
\caption{\label{typeresid}CMB data residuals plotted against
instrument type. The $\chi^2$ for the fit of the line to the data is
$173.9$ with a $\chi^2$ per degree of freedom of 1.00 for the 174
degrees of freedom. The probability of finding a line that better fits
the data is 51\% so the best-fit line improves the fit beyond that of
the zero-line only very slightly.  Thus, there is
no evidence for a trend in this regression plot.}
\end{figure}

\begin{figure}
\plotone{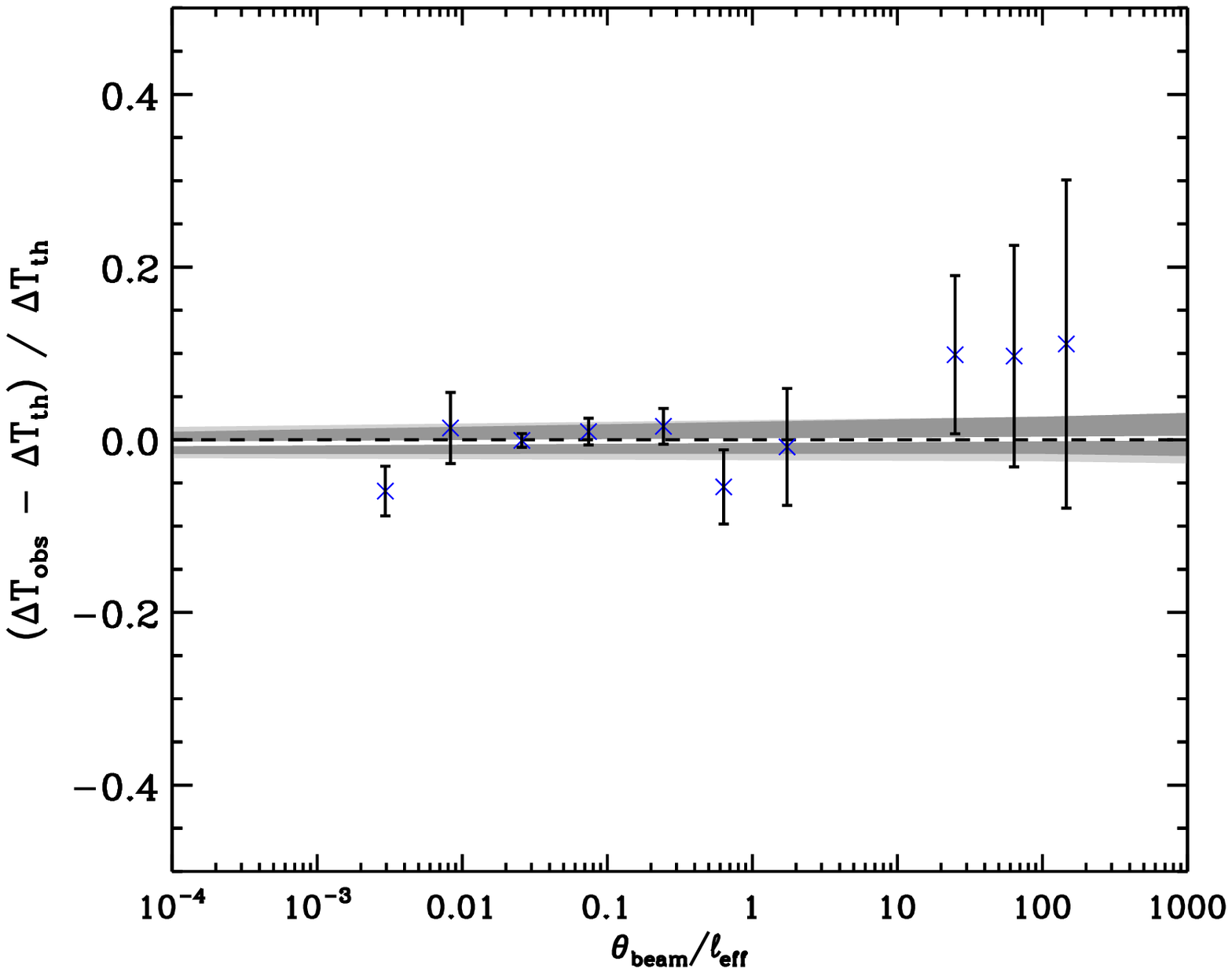}
\caption{\label{resresid}CMB data residuals plotted against
$\theta_{beam}/\ell_{eff}$.  The $x$-axis is logarithmic so as to best
display the data and the residuals are examined for a linear trend
with respect to the logarithmic resolution axis. The $\chi^2$ for the
fit of the line to the data is $173.5$ with a $\chi^2$ per degree of
freedom of 1.00 for the 174 degrees of freedom. The probability of
finding a line that better fits the data is 50\% so the best-fit line
improves the fit beyond that of the zero-line only very slightly.
Thus, there is no evidence for a trend in this
regression plot.}
\end{figure}

\begin{figure}
\plotone{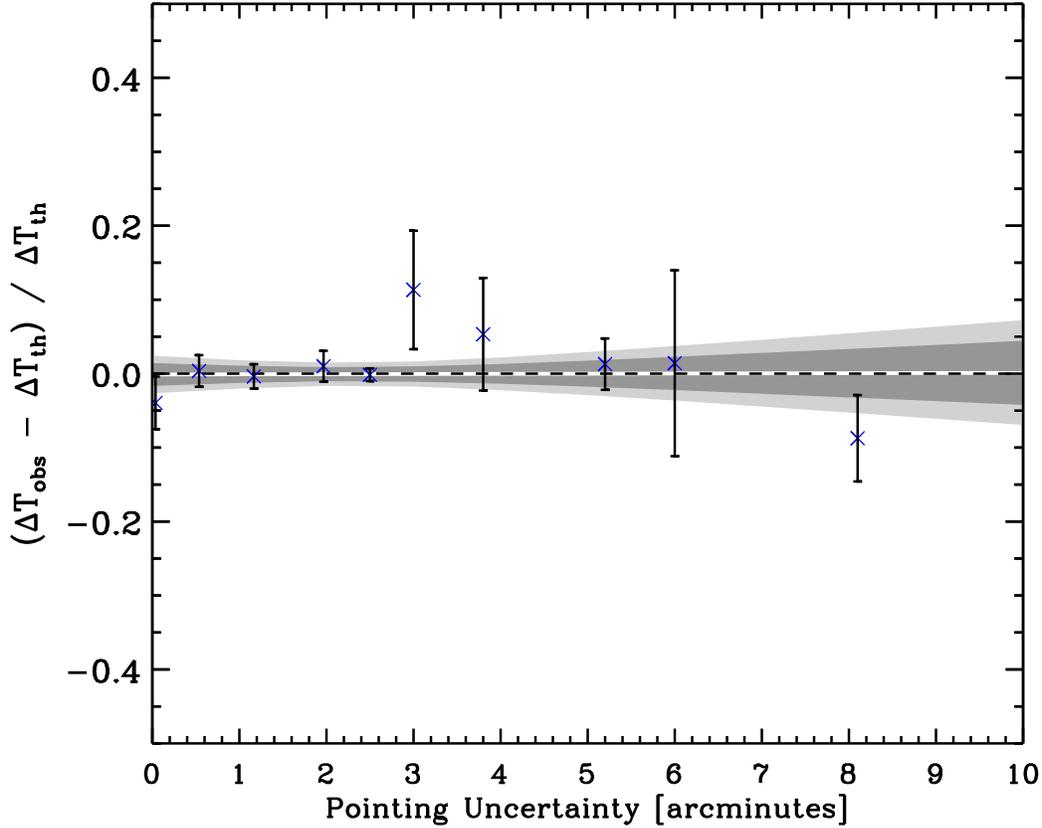}
\caption{\label{pointresid}CMB data residuals plotted against pointing
uncertainty. Six experiments do not quote pointing uncertainties so
are omitted from this analysis. With the 6 experiments omitted, there
are 170 degrees of freedom.  The $\chi^2$ for the zero-line is 172.5
and the probability of finding a line that provides a better fit to
the data is 57\%. The $\chi^2$ for best-fitting line to the residual
data is $172.5$ with a $\chi^2$ per degree of freedom of 1.01 for 170
degrees of freedom. The probability of finding a line that better fits
the data is 57\% so the best-fit line does not improve the fit beyond
that of the zero-line.  Thus, there is no
evidence for a trend in this regression plot. Although, limiting the
analysis to the 5 points with the largest uncertainties would indicate
a trend, suggesting that the largest uncertainties may have been
underestimated.}
\end{figure}

\begin{figure}
\plotone{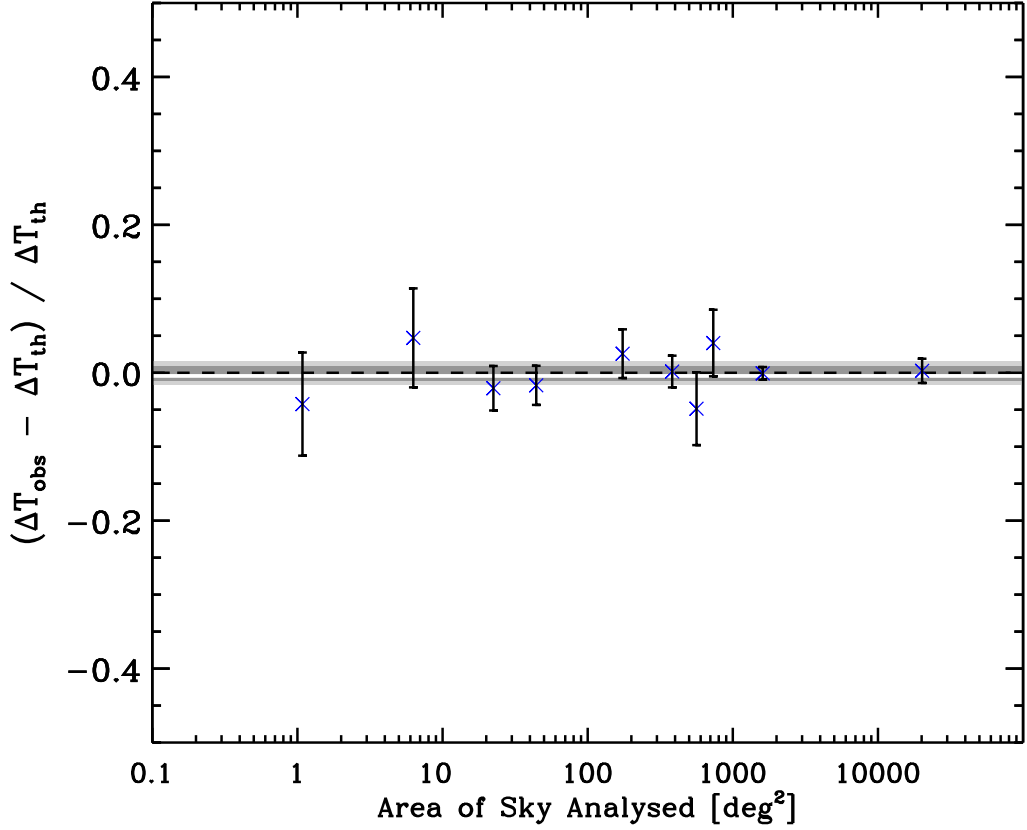}
\caption{\label{arearesid}CMB data residuals plotted against area of
sky.  The area of sky axis is logarithmic so as to best display the
data and the residuals are examined for a linear trend with respect to
the logarithmic $x$-axis.  The $\chi^2$ for the fit of the line
to the data is $172.7$ with a $\chi^2$ per degree of freedom of 0.99
for the 174 degrees of freedom. The probability of finding a line that
better fits the data is 49\% so the best-fit line is a slightly better
fit to the data than the zero-line.  The zero-line is within the 68\%
confidence region for the best-fit line so there
is no evidence for a significant trend in this regression plot.}
\end{figure}

\begin{figure}
\plotone{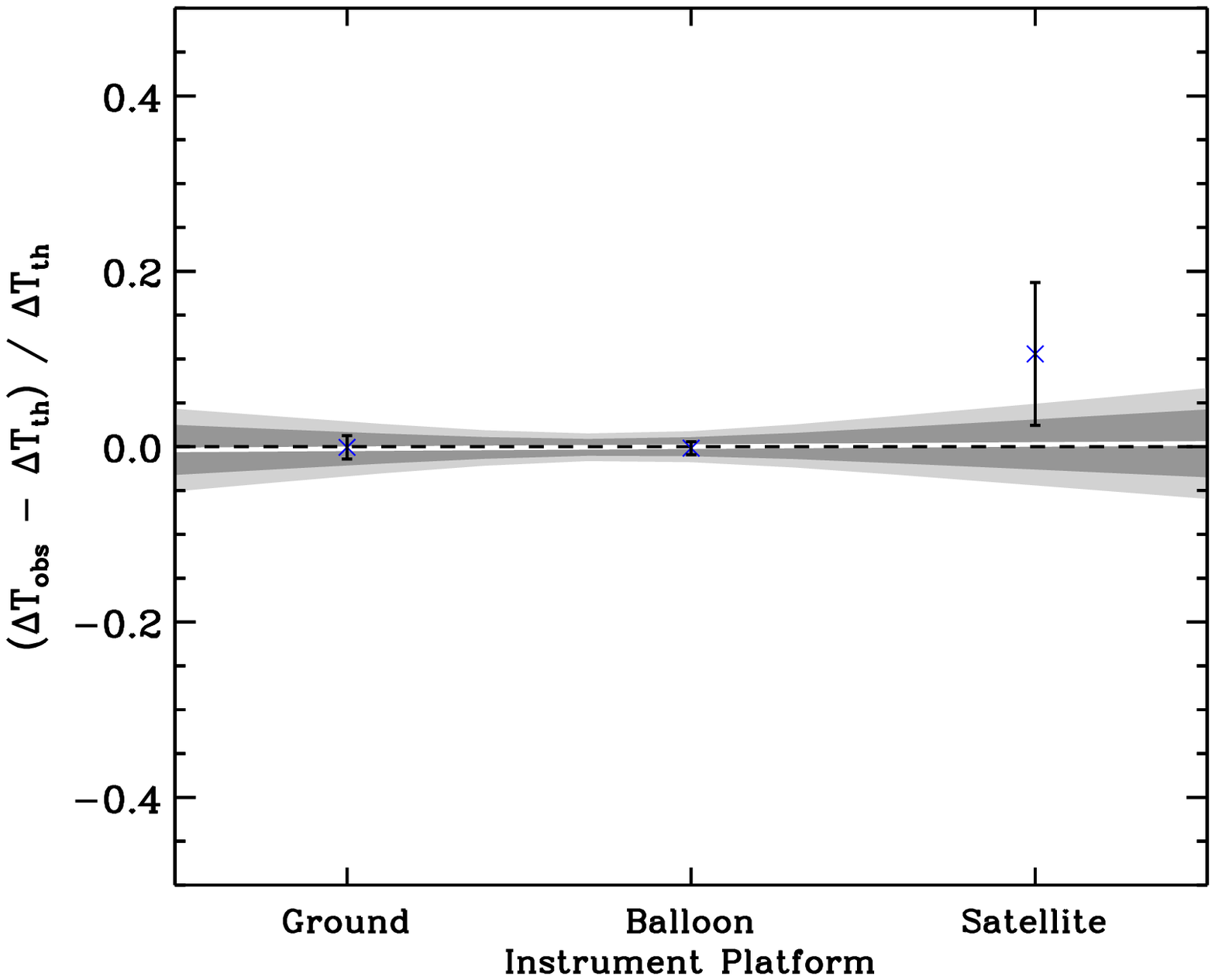}
\caption{\label{platresid}CMB data residuals plotted against
instrument platform. The $\chi^2$ for the fit of the line to the data
is $174.2$ with a $\chi^2$ per degree of freedom of 1.00 for the 174
degrees of freedom. The probability of finding a line that better fits
the data is 52\% so the best-fit line does not improve the fit beyond
that of the zero-line.  Thus, there is no
evidence for a trend in this regression plot.}
\end{figure}

\begin{figure}
\plotone{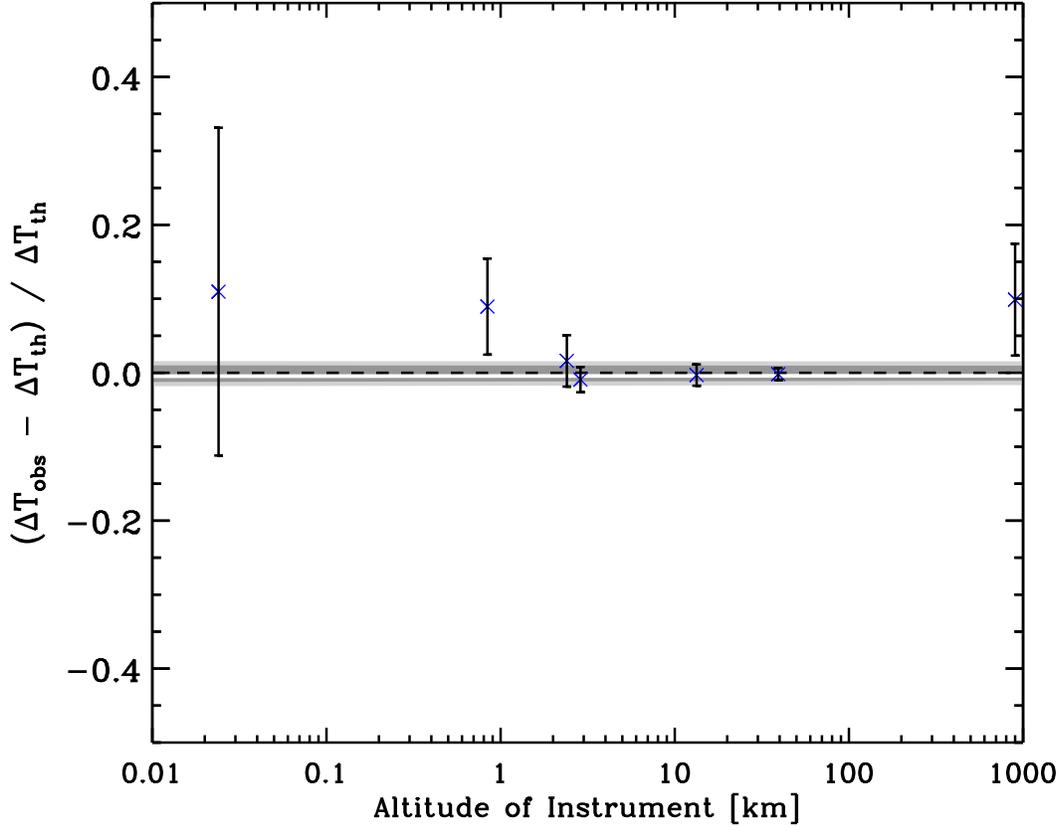}
\caption{\label{altresid}CMB data residuals plotted against instrument
altitude. The $\chi^2$ for the fit of the line to the data is $172.6$
with a $\chi^2$ per degree of freedom of 0.99 for the 174 degrees of
freedom. The probability of finding a line that better fits the data
is 48\% so the best-fit line is a slightly better fit to the data than
the zero-line.  The zero-line is within the 68\% confidence region for
the best-fit line so there is no evidence for a
significant trend in this regression plot.}
\end{figure}

\begin{figure}
\plotone{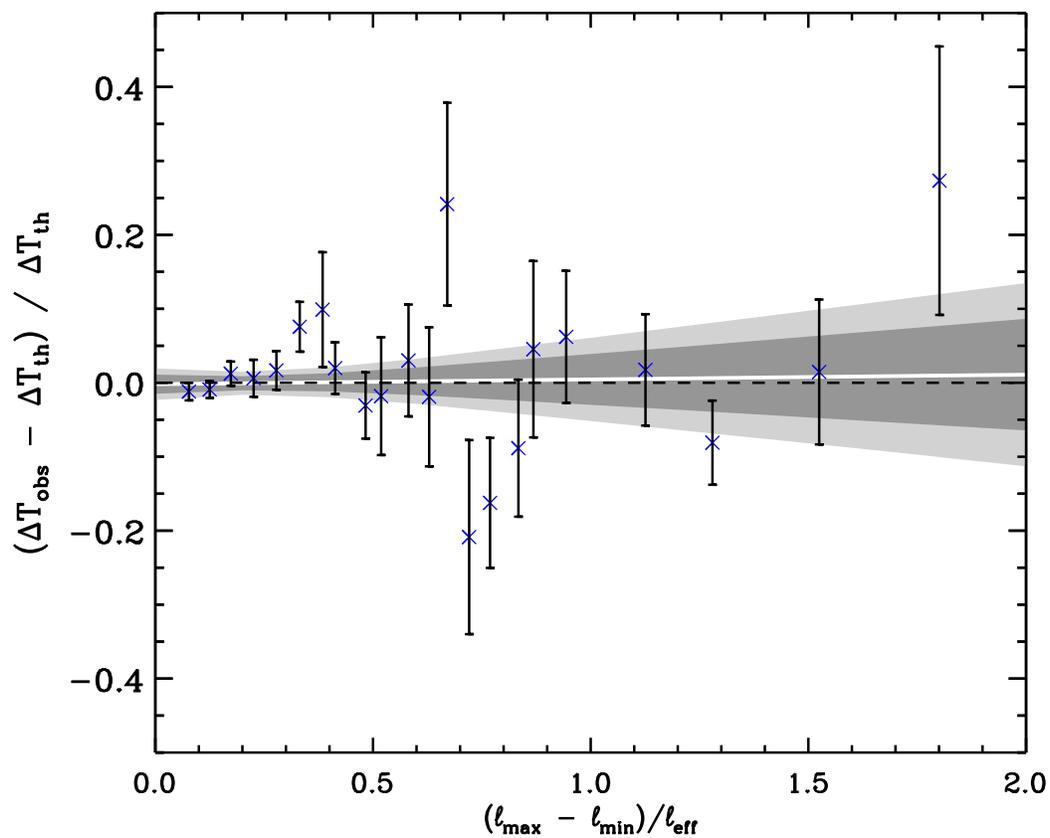} 
\caption{\label{dlresid}CMB data residuals plotted against width of
$\ell$-space filter function $\Delta\ell/\ell$. The $\chi^2$ for the
fit of the line to the data is $174.1$ with a $\chi^2$ per degree of
freedom of 1.00 for the 174 degrees of freedom. The probability of
finding a line that better fits the data is 52\% so the best-fit line
does not improve the fit beyond that of the zero-line.  Thus, there is
no evidence for a trend in this regression plot.}
\end{figure}

\begin{figure}
\plotone{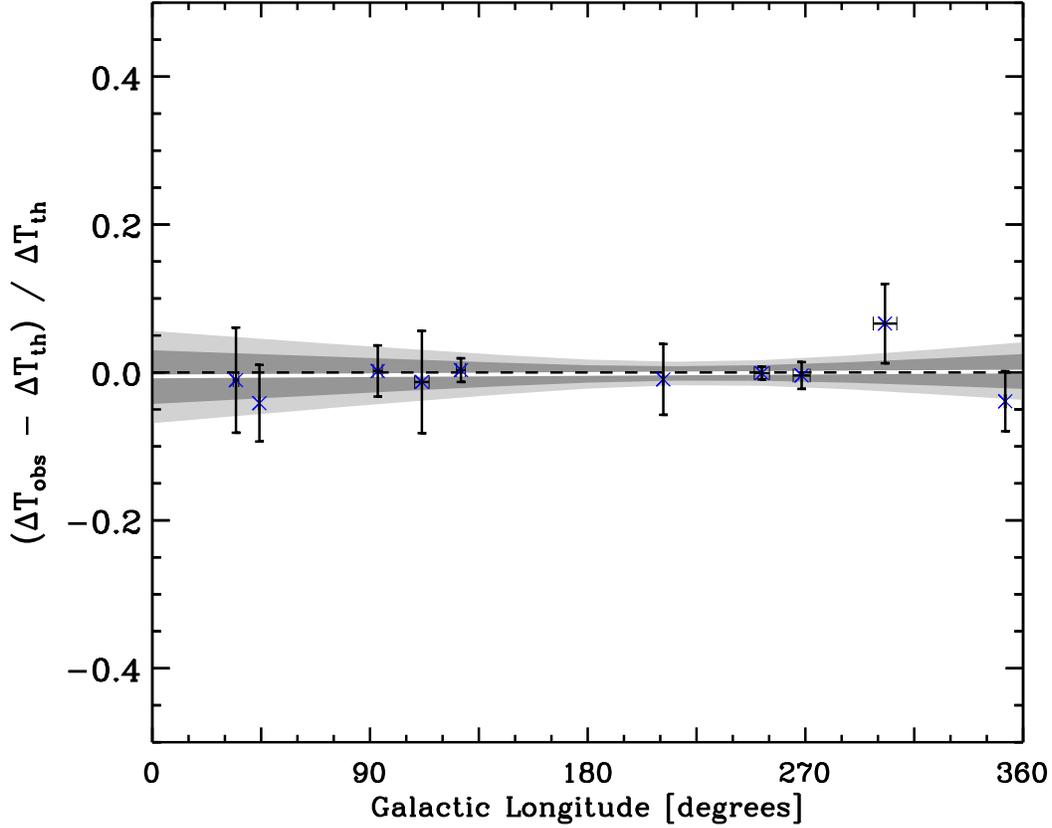} 
\caption{\label{gallongresid}CMB data residuals plotted against
galactic longitude.  The VSA results quoted in the literature are the
combined detections from 3 separate fields observed at various
galactic longitudes.  Information is not given in the literature to
enable the contributions from the different fields to be separated, so
the VSA experiment is omitted from this analysis. With the VSA
experiment omitted, there are 159 degrees of freedom, the $\chi^2$ for
the zero-line is 162.5 and the probability of finding a line that
provides a better fit to the data is 59\%. The $\chi^2$ for the fit of
the line to the data is $162.4$ with a $\chi^2$ per degree of freedom
of 0.98 for the 159 degrees of freedom. The probability of finding a
line that better fits the data is 59\% so the best-fit line does not
improve the fit beyond that of the zero-line.  Thus, there is no
evidence for a trend in this regression plot.}
\end{figure}

\begin{figure}
\plotone{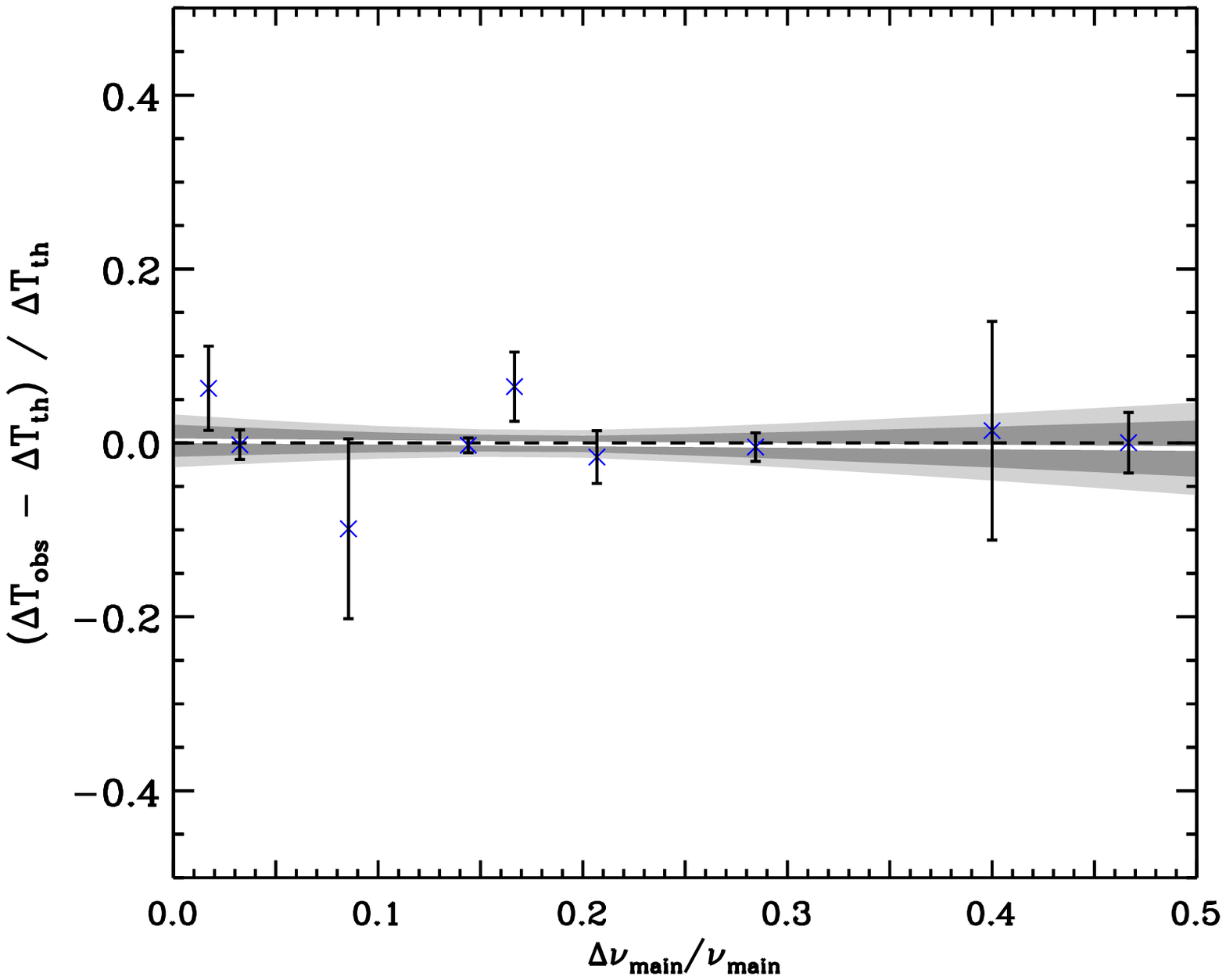} 
\caption{\label{dnuresid}CMB data residuals plotted against bandwidth
$\Delta\nu_{\rm main}/\nu_{\rm main}$. The $\chi^2$ for the fit of the
line to the data is $174.1$ with a $\chi^2$ per degree of freedom of
1.00 for the 174 degrees of freedom. The probability of finding a line
that better fits the data is 52\% so the best-fit line does not
improve the fit beyond that of the zero-line.  Thus, there is no
evidence for a trend in this regression plot.}
\end{figure}

\begin{figure}
\plotone{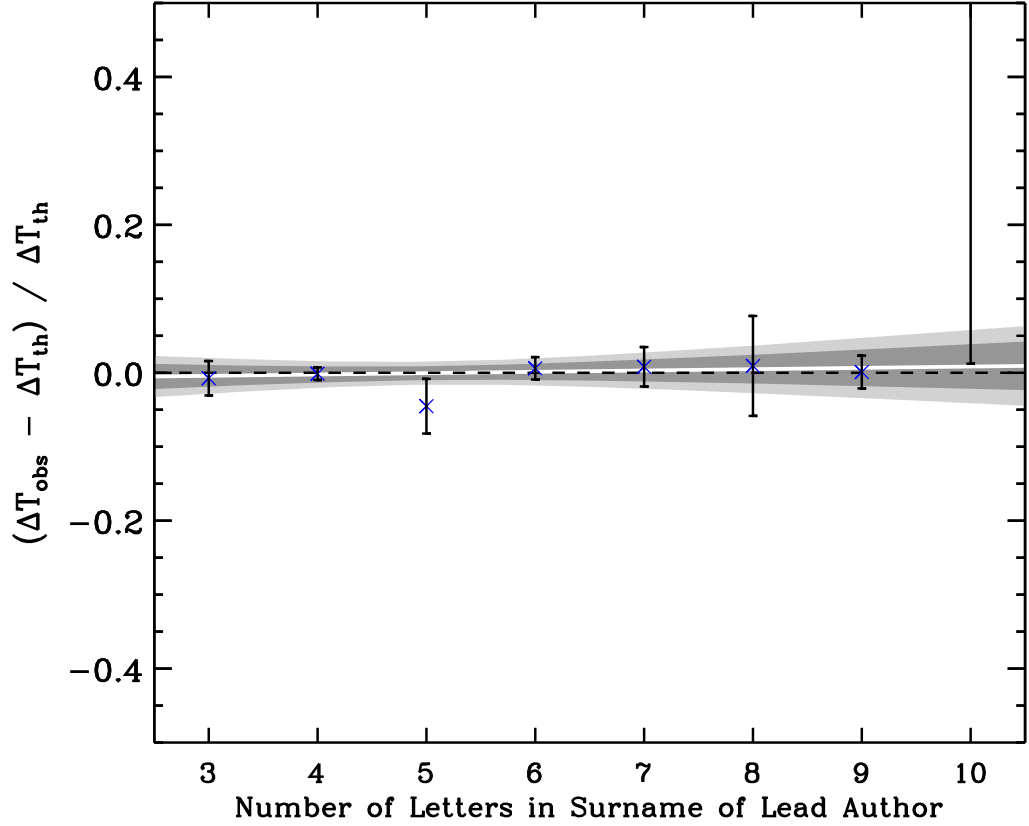}
\caption{\label{nletresid}CMB data residuals plotted against the
number of letters in the surname of the lead author of the published
band-power estimates given in Table~\ref{obsdat}. The $\chi^2$ for the
fit of the line to the data is $174.0$ with a $\chi^2$ per degree of
freedom of 1.00 for the 174 degrees of freedom. The probability of
finding a line that better fits the data is 51\% so the best-fit line
improves the fit beyond that of the zero-line only very slightly.
Therefore, as expected, there is no evidence of a significant trend in
this plot.}
\end{figure}

\begin{figure}
\plotone{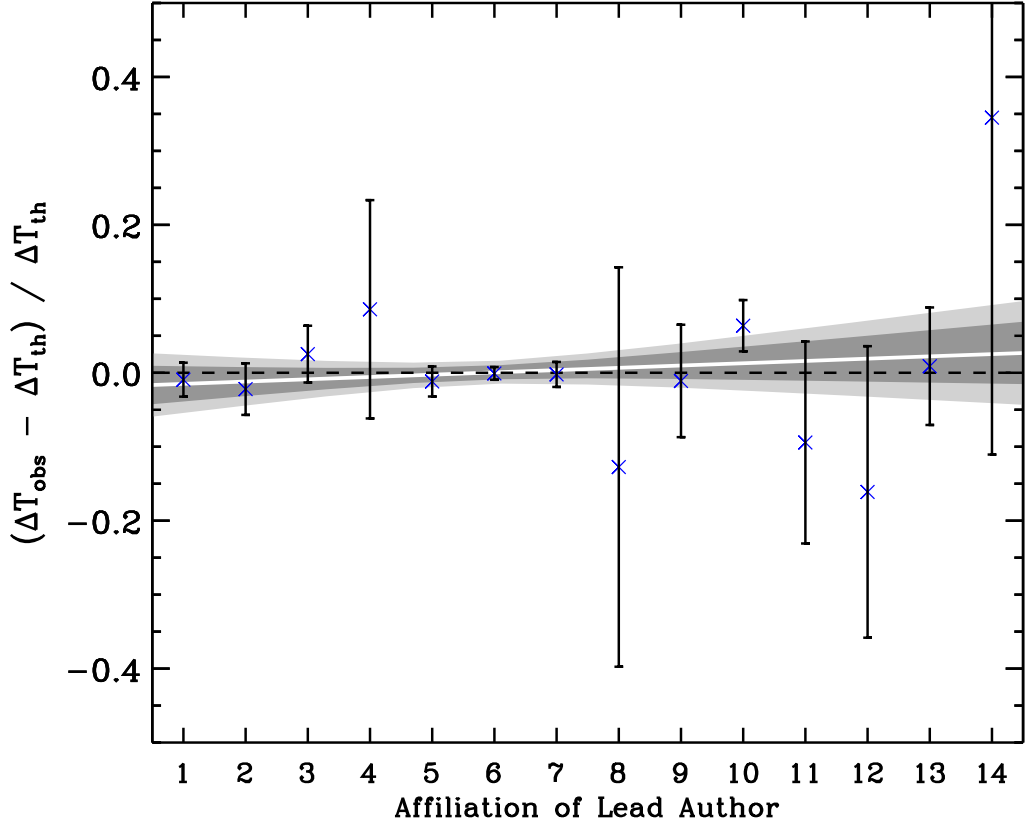}
\caption{\label{affilresid}CMB data residuals plotted against the
affiliation of the lead author of the published band-power estimates
given in Table~\ref{obsdat} (1:Berkeley, 2:CalTech, 3:Cambridge,
4:Carnegie Mellon, 5:Chicago, 6:Cleveland, 7:France, 8:Manchester,
9:Massachusetts, 10:Princeton, 11:Rome, 12:Santa Barbara, 13:Spain,
14:UBC). The $\chi^2$ for the fit of the line to the data is $173.2$
with a $\chi^2$ per degree of freedom of 1.00 for the 174 degrees of
freedom. The probability of finding a line that better fits the data
is 50\% so the best-fit line improves the fit beyond that of the
zero-line only very slightly.  Therefore, as expected, there is no
evidence of a significant trend in this plot.}
\end{figure}

\begin{figure}
\plotone{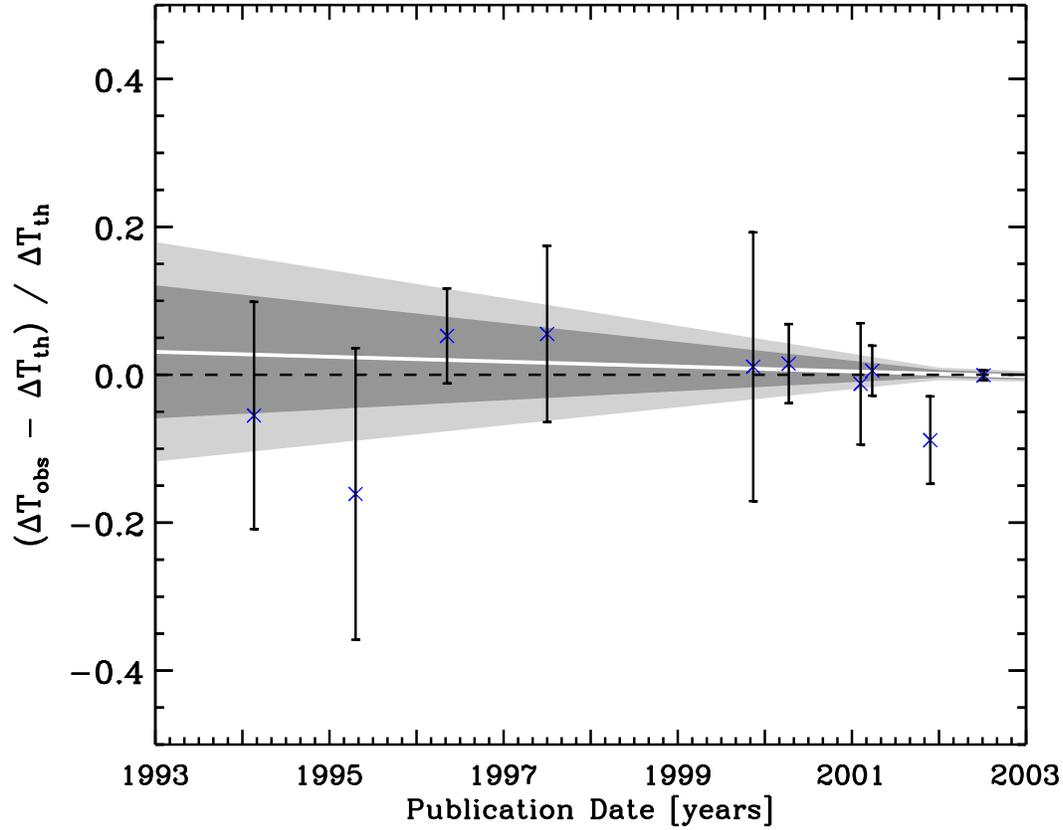}
\caption{\label{pdresid}CMB data residuals plotted against the
publication date of the band-power estimates given in
Table~\ref{obsdat}. The $\chi^2$ for the fit of the line to the data
is $173.9$ with a $\chi^2$ per degree of freedom of 1.00 for the 174
degrees of freedom. The probability of finding a line that better fits
the data is 51\% so the best-fit line improves the fit beyond that of the
zero-line only very slightly.  Therefore, as expected, there is no
evidence of a significant trend in this plot.}
\end{figure}

The ever increasing number of CMB anisotropies has made data plots
such as Figure~\ref{clplotcol} difficult to interpret.  The solution
is to compress the data in some way.  Many of the more recent analyses
have chosen to concentrate on the data from just one or two
experiments, often the most recently released.  However, this not only
neglects potentially useful information, but can also unwittingly give
more weight to particular observations that may suffer from associated
systematic effects. We therefore choose to analyse all the available
data.

One way to compress the data is to average them together into single
band-power bins in $\ell$-space.  Such an approach has been taken by a
number of authors \citep[e.g.][]{knoxpage00,jaffeetal01,wangetal02a}.
Providing that the uncertainty in the data is Gaussian and
correlations between detections are treated appropriately, narrow band
power bins can be chosen that will retain all cosmological
information.

Band-power measurements from independent observations that overlap in
the sky will be correlated to some extent.  Such correlations can only
be treated by jointly analysing the combined overlapping maps to
extract band power estimates that are uncorrelated or have explicitly
defined correlation matrices.  This process of data compression will
wash out any systematics associated with a particular data set, so
data consistency checks are vital before this stage.  If the map data
is unavailable, the crude assumption that independent observations are
uncorrelated in space must be made.  This assumption is made in
likelihood analyses performed on the full power spectrum data set and,
since inclusion of these correlations would reduce the degrees of
freedom of an analysis, the goodness-of-fit of a particular model to
the data is better than it should be.

Some observational groups publish matrices encoding the correlations
of their individual band-power measurements.  To some extent, the
calibration uncertainties of experiments that calibrate using the same
source are also correlated. \citet{bondetal00} describe a data binning
technique that takes a lognormal noise distribution that is
approximately Gaussian and incorporates the correlation weight
matrices of individual experiments. \citet{wangetal02a} detail a
method to treat partial correlations of calibration uncertainties.
Both are useful to produce statistically meaningful data bins.

Data binning averages out any evidence for discrepancies between
independent observations and, in practice, data uncertainties are
rarely Gaussian and the information required to treat correlated data
is not always available.  So although data binning is useful for
visualisation purposes, statistical analyses of the binned
observations will generally give different results from those
performed on the raw data.

The statistical analyses detailed in this paper are performed on the
published CMB band power measurements given in Table~\ref{obsdat}.
Binned data plots are presented purely to aid the interpretation of
results. Therefore each calibrated and, for BOOMERanG98 and MAXIMA1,
beam corrected observational data point is binned assuming it to be
entirely uncorrelated.  Bin widths must be carefully chosen so that
important features of the data are not smoothed out, especially in
regions of extreme curvature.  For example, in the case of the power
spectrum, an unwisely chosen bin that spans an acoustic maximum will
average out the power in the bin to produce a binned data point that
misleadingly assigns less power to the peak.

The contribution from the $i$th observational measurement
($x_i\pm\sigma_{x, \, i}$, $y_i\pm\sigma_{y,\, i}$) to a binned point
($x_b\pm\sigma_{x, \, b}$, $y_b\pm\sigma_{y,\, b}$) is weighted by the
square inverse of its variance,
\begin{eqnarray}
x_b \, \pm \, \sigma_{x, \, b} \, = \, \frac{\sum_i x_i \, \sigma_{x, \, i}^{-2}}{\sum_i\sigma_{x, \, i}^{-2}} \, \pm \, \sqrt{\frac{1}{\sum_i\sigma_{x, \, i}^{-2}}} \,, \\
y_b \, \pm \, \sigma_{y, \, b}\,  = \, \frac{\sum_i y_i \, \sigma_{y, \, i}^{-2}}{\sum_i\sigma_{y, \, i}^{-2}} \, \pm \, \sqrt{\frac{1}{\sum_i\sigma_{y, \, i}^{-2}}} \,.
\end{eqnarray}
If quoted error bars are uneven, a first guess for the binned data
point is obtained by averaging the uncertainties.  A more accurate
estimate can then be converged upon by iterating over the binning
routine, inserting the positive variance for measurements that are
below the bin averaged point and negative variance for those that are
above.

All the data from a particular experiment will be measured using the
same instrument and therefore can be binned together for the purpose
of visualising any trends in the data residuals with respect to the
instrument design. When data from the same experiment is placed in the
same bin, the variance of the resultant binned data point can be
easily adjusted to account for any correlated calibration uncertainty
associated with the observational data. A degree of freedom is reduced
for each independent calibration uncertainty. This effectively
tightens the constraints on the binned data.

For example, if $n$ calibrated data points in a bin have equal
variance $\sigma_y$ and an entirely correlated calibration
uncertainty, they share $n-1$ degrees of freedom and their
contribution to the variance of the binned data point is then
$\sigma_y/\sqrt{n-1}$. When each of the $n$ data points have different
variances, their contribution to the uncertainty in the binned data
point is given by,
\begin{equation}
\sigma_{y, \, b} \, = \, \frac{1}{\sqrt{\sum_j\sigma_{y, \, j}^{-2} - \left(n \, / \, \sum_j\sigma_{y, \, j}\right)^2}} \,.
\end{equation}
This method of binning is employed when appropriate to produce the
plotted residual data bins.

\clearpage

\clearpage

\clearpage

\clearpage

\clearpage

\clearpage

\clearpage

\clearpage

\clearpage

\clearpage

\clearpage

\clearpage

\clearpage

\clearpage

\clearpage

\clearpage

\clearpage

\clearpage

\clearpage

\clearpage

\clearpage
\begin{deluxetable}{lccccccccccc} 
\rotate
\tabletypesize{\tiny}
\tablecolumns{12} 
\tablewidth{0pc}
\setlength{\tabcolsep}{0.04in}
\tablecaption{\label{concord} 1-$\sigma$ cosmological parameter constraints from 5 independent analyses.}
\tablehead{ 
\colhead{}    &  \multicolumn{2}{c}{\cite{efstetal02}} & \multicolumn{2}{c}{\cite{wangetal02a}}  & \multicolumn{2}{c}{\cite{sievers02}} & \multicolumn{2}{c}{\cite{lewisbridle02}} & \multicolumn{2}{c}{\cite{wangetal02b}} & $\Lambda$CDM concordance \\
\colhead{} & \colhead{CMB alone}   & \colhead{+2dFGRS+BBN} & \colhead{CMB alone} & \colhead{+PSC$z$+HKP}    & \colhead{CMB alone}   & \colhead{+priors$^a$} & \colhead{CMB+priors$^b$} &\colhead{+2dFGRS} & \colhead{CMB+flat prior}   & \colhead{+2dFGRS} &\colhead{} }
\startdata
$\Omega_k$       &$-0.04^{+.05}_{-.32}$  &$-0.013^{+.027}_{-.019}$&$-0.05^{+.05}_{-.17}$   &$0.02^{+.03}_{-.04}$   &$-0.05\pm .05$         & (0.00)                & (0.00)           & (0.00)           & (0.00) 		& (0.00)          & 0.0    \\
$\Omega_\Lambda$ &$0.43^{+.23}$          &$0.73\pm .04$           &$0.57^{+.16}_{-.23}$    &$0.65^{+.05}_{-.09}$    &$0.54^{+.12}_{-.13}$   &$0.70^{+.02}_{-.03}$   & $0.72\pm 0.06$   & $0.71 \pm 0.04$  &$0.71 \pm 0.11$   & $0.72 \pm 0.09$ &0.7 \\
$\omega_b$       &$0.020^{+.013}_{-.002}$&$0.020\pm .001$         &$0.021^{+.005}_{-.003}$ &$0.020^{+.005}_{-.002}$ &$0.023\pm .003$        &$0.024^{+.002}_{-.003}$& $0.022 \pm .001$ & $0.022 \pm .001$ & 0.023$\pm$.003 	& 0.024$\pm$.003 &0.02\\
$\omega_c$       &$0.13^{+.03}_{-.05}$   &$0.10^{+.02}_{-.01}$    &-                       &-                       & $0.13^{+.03}_{-.02}$  &$0.12^{+.01}_{-.01}$   & -                & -                & 0.112$\pm$.014 	& 0.115$\pm$.013  &0.12\\
$\omega_d$       &-                      &-                       &$0.10^{+.02}_{-.03}$    &$0.12^{+.03}_{-.02}$    &-                      &-                      & 0.099$\pm$.014   & 0.106$\pm$.010   & - 		& -  &0.12\\
$f_{\nu}$	    &-                      &-                       &$0.08^{+0.30}$          &$0.04^{+.08}$           &-                      &-                      &$<0.10$           &$<0.04$           & - 		& -       &0             \\
$n_s$            &$0.96^{+.27}_{-.04}$ &$1.04^{+.06}_{-.05}$      &$0.91^{+.08}_{-.05}$    &$0.93^{+.06}_{-.05}$    &$1.02^{+.06}_{-.07}$   &$1.04^{+.05}_{-.06}$   & 1.02$\pm$.05     &1.03$\pm$.05      &0.99$\pm$.06 	& 0.99$\pm$.04 &1.0 \\
$\tau$           &$<0.25$                &$<0.25$                 &$<0.09$                 &$<0.14$                 &$0.16^{+.18}_{-.13}$   &$0.13^{+.13}_{-.10}$   & -                & -                & $0.04^{+0.06}$		&$0.06 \pm .03$ &0  \\
$\Omega_m h$     &-                      &$0.19 \pm .02$          &-                       &-                       &-                      &-                      & 0.18$\pm$.03     &0.19$\pm$.02      &- 		&-                 &0.2\\
$h$              &-                      &$0.66^{+.09}_{-.03}$    &$0.42^{+.12}_{-.09}$    &$0.71^{+.06}_{-.06}$    &$0.55^{+.09}_{-.09}$   &$0.69^{+.02}_{-.02}$   & 0.67$\pm$.05     & 0.66$\pm$.03     & 0.71$\pm$.13 	& 0.73$\pm$.11 & 0.68\\
\enddata 
\tablenotetext{a}{The priors used in this analysis are a flat prior $\Omega_k=0$ in accordance with the predictions of the simplest inflationary scenarios, a large scale structure prior that involves a constraint on the amplitude $\sigma_8^2$ and the shape of the matter power spectrum, the HKP prior for $h$ and the SNIa priors.}
\tablenotetext{b}{The priors used in this analysis are a flat prior $\Omega_k=0$ in accordance with the predictions of the simplest inflationary scenarios, the BBN prior for $\omega_b$, the HKP prior for $h$ and the SNIa priors.}
\end{deluxetable}
\clearpage
\begin{deluxetable}{lccccccc}
\tabletypesize{\tiny}
\tablecolumns{8} 
\tablewidth{0pc} 
\tablecaption{\label{obsdat} The current compilation of CMB observational data from $\ell=2$ to $\ell=2000$.} 
\tablehead{ 
\colhead{Experiment} & \colhead{Ref.}   & \colhead{$\ell_{{\rm eff}}$}    & \colhead{$\ell_{{\rm min}}$} & 
\colhead{$\ell_{{\rm max}}$}    & \colhead{$\delta T_{\ell_{\rm eff}}^{\rm obs}\pm \sigma^{\rm obs}$}   & \colhead{$\sigma_{c}^{a}$}    & \colhead{ Publication Date}\\
\colhead{} & \colhead{}   & \colhead{}    & \colhead{} & 
\colhead{}    & \colhead{ $(\mu$K)}   & \colhead{(\%)}    & \colhead{(yrs)}}
\startdata 
       ACBAR &  [1] &  187.0 &   75.0 &  300.0 & $ 82.3_{- 8.5}^{+ 7.7}$ & 10.0 & 2002.9 \\
       ACBAR &  [1] &  389.0 &  307.0 &  459.0 & $ 53.6_{- 6.0}^{+ 5.4}$ & 10.0 & 2002.9 \\
       ACBAR &  [1] &  536.0 &  462.0 &  602.0 & $ 52.1_{- 5.0}^{+ 4.6}$ & 10.0 & 2002.9 \\
       ACBAR &  [1] &  678.0 &  615.0 &  744.0 & $ 47.1_{- 4.0}^{+ 3.7}$ & 10.0 & 2002.9 \\
       ACBAR &  [1] &  842.0 &  751.0 &  928.0 & $ 48.0_{- 3.9}^{+ 3.6}$ & 10.0 & 2002.9 \\
       ACBAR &  [1] &  986.0 &  921.0 & 1048.0 & $ 28.2_{- 2.9}^{+ 2.6}$ & 10.0 & 2002.9 \\
       ACBAR &  [1] & 1128.0 & 1040.0 & 1214.0 & $ 36.1_{- 3.0}^{+ 2.8}$ & 10.0 & 2002.9 \\
       ACBAR &  [1] & 1279.0 & 1207.0 & 1352.0 & $ 24.1_{- 2.9}^{+ 2.6}$ & 10.0 & 2002.9 \\
       ACBAR &  [1] & 1426.0 & 1338.0 & 1513.0 & $ 25.1_{- 2.8}^{+ 2.5}$ & 10.0 & 2002.9 \\
       ACBAR &  [1] & 1580.0 & 1510.0 & 1649.0 & $ 18.7_{- 3.2}^{+ 2.7}$ & 10.0 & 2002.9 \\
       ACBAR &  [1] & 1716.0 & 1648.0 & 1785.0 & $ 15.7_{- 3.5}^{+ 2.9}$ & 10.0 & 2002.9 \\
       ACBAR &  [1] & 1866.0 & 1782.0 & 1953.0 & $ 19.0_{- 3.9}^{+ 3.2}$ & 10.0 & 2002.9 \\
    ACME-MAX &  [2] &  139.0 &   72.0 &  247.0 & $ 49.4_{- 7.8}^{+ 7.8}$ & -  &1996.7 \\
   ACME-SP91 &  [3] &   61.0 &   30.0 &  102.0 & $ 30.3_{- 8.9}^{+ 5.6}$ & -  &1995.3 \\
   ACME-SP94 &  [3] &   61.0 &   30.0 &  102.0 & $ 36.3_{-14.1}^{+ 6.9}$ & -  &1995.3 \\
    ARCHEOPS &  [4] &   18.5 &   15.0 &   22.0 & $ 28.1_{-12.2}^{+ 8.3}$ &  7.0 & 2002.8 \\
    ARCHEOPS &  [4] &   28.5 &   22.0 &   35.0 & $ 30.6_{- 4.0}^{+ 3.6}$ &  7.0 & 2002.8 \\
    ARCHEOPS &  [4] &   40.0 &   35.0 &   45.0 & $ 34.6_{- 4.0}^{+ 3.6}$ &  7.0 & 2002.8 \\
    ARCHEOPS &  [4] &   52.5 &   45.0 &   60.0 & $ 30.2_{- 4.0}^{+ 3.5}$ &  7.0 & 2002.8 \\
    ARCHEOPS &  [4] &   70.0 &   60.0 &   80.0 & $ 39.9_{- 2.9}^{+ 2.7}$ &  7.0 & 2002.8 \\
    ARCHEOPS &  [4] &   87.5 &   80.0 &   95.0 & $ 44.2_{- 3.3}^{+ 3.1}$ &  7.0 & 2002.8 \\
    ARCHEOPS &  [4] &  102.5 &   95.0 &  110.0 & $ 51.2_{- 3.3}^{+ 3.1}$ &  7.0 & 2002.8 \\
    ARCHEOPS &  [4] &  117.5 &  110.0 &  125.0 & $ 51.8_{- 3.6}^{+ 3.4}$ &  7.0 & 2002.8 \\
    ARCHEOPS &  [4] &  135.0 &  125.0 &  145.0 & $ 58.8_{- 3.1}^{+ 3.0}$ &  7.0 & 2002.8 \\
    ARCHEOPS &  [4] &  155.0 &  145.0 &  165.0 & $ 60.7_{- 3.4}^{+ 3.2}$ &  7.0 & 2002.8 \\
    ARCHEOPS &  [4] &  175.0 &  165.0 &  185.0 & $ 67.7_{- 3.5}^{+ 3.3}$ &  7.0 & 2002.8 \\
    ARCHEOPS &  [4] &  197.5 &  185.0 &  210.0 & $ 69.3_{- 3.5}^{+ 3.3}$ &  7.0 & 2002.8 \\
    ARCHEOPS &  [4] &  225.0 &  210.0 &  240.0 & $ 67.5_{- 3.6}^{+ 3.4}$ &  7.0 & 2002.8 \\
    ARCHEOPS &  [4] &  257.5 &  240.0 &  275.0 & $ 71.1_{- 3.5}^{+ 3.4}$ &  7.0 & 2002.8 \\
    ARCHEOPS &  [4] &  292.5 &  275.0 &  310.0 & $ 57.5_{- 5.1}^{+ 4.7}$ &  7.0 & 2002.8 \\
    ARCHEOPS &  [4] &  330.0 &  310.0 &  350.0 & $ 51.3_{- 4.8}^{+ 4.4}$ &  7.0 & 2002.8 \\
       ARGO1 &  [5] &   95.0 &   51.0 &  173.0 & $ 39.1_{- 8.7}^{+ 8.7}$ & -  &1994.1 \\
       ARGO2 &  [6] &   95.0 &   51.0 &  173.0 & $ 46.8_{- 9.5}^{+12.1}$ & -  &1996.4 \\
         BAM &  [7] &   74.0 &   27.0 &  156.0 & $ 55.6_{-31.6}^{+18.8}$ & -  &1997.2 \\
      BOOM97 &  [8] &   58.0 &   25.0 &   75.0 & $ 29.0_{-11.0}^{+13.0}$ &  8.0 & 2000.4 \\
      BOOM97 &  [8] &  102.0 &   76.0 &  125.0 & $ 49.0_{- 9.0}^{+ 9.0}$ &  8.0 & 2000.4 \\
      BOOM97 &  [8] &  153.0 &  126.0 &  175.0 & $ 67.0_{- 9.0}^{+10.0}$ &  8.0 & 2000.4 \\
      BOOM97 &  [8] &  204.0 &  176.0 &  225.0 & $ 72.0_{-10.0}^{+10.0}$ &  8.0 & 2000.4 \\
      BOOM97 &  [8] &  255.0 &  226.0 &  275.0 & $ 61.0_{-12.0}^{+11.0}$ &  8.0 & 2000.4 \\
      BOOM97 &  [8] &  305.0 &  276.0 &  325.0 & $ 55.0_{-15.0}^{+14.0}$ &  8.0 & 2000.4 \\
      BOOM97 &  [8] &  403.0 &  326.0 &  475.0 & $ 32.0_{-22.0}^{+13.0}$ &  8.0 & 2000.4 \\
      BOOM98 &  [9] &   50.5 &   26.0 &   75.0 & $ 37.7_{- 4.4}^{+ 3.9}$ & 10.0 & 2002.9 \\
      BOOM98 &  [9] &  100.5 &   76.0 &  125.0 & $ 51.1_{- 2.8}^{+ 2.7}$ & 10.0 & 2002.9 \\
      BOOM98 &  [9] &  150.5 &  126.0 &  175.0 & $ 69.4_{- 2.8}^{+ 2.7}$ & 10.0 & 2002.9 \\
      BOOM98 &  [9] &  200.5 &  176.0 &  225.0 & $ 71.7_{- 2.5}^{+ 2.4}$ & 10.0 & 2002.9 \\
      BOOM98 &  [9] &  250.5 &  226.0 &  275.0 & $ 73.2_{- 2.2}^{+ 2.2}$ & 10.0 & 2002.9 \\
      BOOM98 &  [9] &  300.5 &  276.0 &  325.0 & $ 62.9_{- 1.8}^{+ 1.7}$ & 10.0 & 2002.9 \\
      BOOM98 &  [9] &  350.5 &  326.0 &  375.0 & $ 49.4_{- 1.4}^{+ 1.4}$ & 10.0 & 2002.9 \\
      BOOM98 &  [9] &  400.5 &  376.0 &  425.0 & $ 42.7_{- 1.2}^{+ 1.2}$ & 10.0 & 2002.9 \\
      BOOM98 &  [9] &  450.5 &  426.0 &  475.0 & $ 45.7_{- 1.3}^{+ 1.3}$ & 10.0 & 2002.9 \\
      BOOM98 &  [9] &  500.5 &  476.0 &  525.0 & $ 49.6_{- 1.4}^{+ 1.3}$ & 10.0 & 2002.9 \\
      BOOM98 &  [9] &  550.5 &  526.0 &  575.0 & $ 49.4_{- 1.4}^{+ 1.3}$ & 10.0 & 2002.9 \\
      BOOM98 &  [9] &  600.5 &  576.0 &  625.0 & $ 47.1_{- 1.4}^{+ 1.4}$ & 10.0 & 2002.9 \\
      BOOM98 &  [9] &  650.5 &  626.0 &  675.0 & $ 44.7_{- 1.5}^{+ 1.5}$ & 10.0 & 2002.9 \\
      BOOM98 &  [9] &  700.5 &  676.0 &  725.0 & $ 46.8_{- 1.7}^{+ 1.6}$ & 10.0 & 2002.9 \\
      BOOM98 &  [9] &  750.5 &  726.0 &  775.0 & $ 44.8_{- 2.0}^{+ 1.9}$ & 10.0 & 2002.9 \\
      BOOM98 &  [9] &  800.5 &  776.0 &  825.0 & $ 50.8_{- 2.2}^{+ 2.1}$ & 10.0 & 2002.9 \\
      BOOM98 &  [9] &  850.5 &  826.0 &  875.0 & $ 47.2_{- 2.7}^{+ 2.5}$ & 10.0 & 2002.9 \\
      BOOM98 &  [9] &  900.5 &  876.0 &  925.0 & $ 47.5_{- 3.2}^{+ 3.0}$ & 10.0 & 2002.9 \\
      BOOM98 &  [9] &  950.5 &  926.0 &  975.0 & $ 34.0_{- 5.3}^{+ 4.6}$ & 10.0 & 2002.9 \\
      BOOM98 &  [9] & 1000.5 &  976.0 & 1025.0 & $ 34.0_{- 7.1}^{+ 5.8}$ & 10.0 & 2002.9 \\
        CAT1 & [10] &  397.0 &  322.0 &  481.0 & $ 50.8_{-15.4}^{+15.4}$ & 10.0 & 1996.3 \\
        CAT1 & [10] &  615.0 &  543.0 &  717.0 & $ 49.0_{-13.6}^{+19.1}$ & 10.0 & 1996.3 \\
        CAT2 & [11] &  397.0 &  322.0 &  481.0 & $ 57.3_{-13.6}^{+10.9}$ & 10.0 & 1999.8 \\
        CAT2 & [11] &  615.0 &  543.0 &  717.0 & $  0.0_{- 0.0}^{+54.6}$ & 10.0 & 1999.8 \\
      CBI-1a & [12] &  603.0 &  437.0 &  783.0 & $ 64.0_{- 9.0}^{+11.0}$ &  5.0 & 2001.2 \\
      CBI-1a & [12] & 1190.0 &  966.0 & 1451.0 & $ 31.0_{- 5.0}^{+ 7.0}$ &  5.0 & 2001.2 \\
      CBI-1b & [12] &  603.0 &  437.0 &  783.0 & $ 52.0_{- 9.0}^{+11.0}$ &  5.0 & 2001.2 \\
      CBI-1b & [12] & 1190.0 &  966.0 & 1451.0 & $ 28.0_{- 7.0}^{+10.0}$ &  5.0 & 2001.2 \\
     CBI-D8h & [13] &  307.0 &    2.0 &  500.0 & $ 80.8_{-31.5}^{+22.3}$ &  5.0 & 2002.4 \\
     CBI-D8h & [13] &  640.0 &  500.0 &  880.0 & $ 56.7_{-13.0}^{+10.5}$ &  5.0 & 2002.4 \\
     CBI-D8h & [13] & 1133.0 &  880.0 & 1445.0 & $ 32.3_{- 6.7}^{+ 5.5}$ &  5.0 & 2002.4 \\
     CBI-D8h & [13] & 1703.0 & 1445.0 & 2010.0 & $ 21.2_{- 7.7}^{+ 5.5}$ &  5.0 & 2002.4 \\
    CBI-D14h & [13] &  307.0 &    2.0 &  500.0 & $ 91.5_{-35.8}^{+25.3}$ &  5.0 & 2002.4 \\
    CBI-D14h & [13] &  640.0 &  500.0 &  880.0 & $ 30.2_{-30.2}^{+13.3}$ &  5.0 & 2002.4 \\
    CBI-D14h & [13] & 1133.0 &  880.0 & 1445.0 & $ 17.6_{-17.6}^{+10.2}$ &  5.0 & 2002.4 \\
    CBI-D14h & [13] & 1703.0 & 1445.0 & 2010.0 & $  0.0_{- 0.0}^{+10.0}$ &  5.0 & 2002.4 \\
    CBI-D20h & [13] &  307.0 &    2.0 &  500.0 & $ 82.6_{-25.3}^{+19.2}$ &  5.0 & 2002.4 \\
    CBI-D20h & [13] &  640.0 &  500.0 &  880.0 & $ 36.4_{-10.0}^{+ 7.8}$ &  5.0 & 2002.4 \\
    CBI-D20h & [13] & 1133.0 &  880.0 & 1445.0 & $ 26.0_{- 5.7}^{+ 4.7}$ &  5.0 & 2002.4 \\
    CBI-D20h & [13] & 1703.0 & 1445.0 & 2010.0 & $ 24.0_{- 6.1}^{+ 4.9}$ &  5.0 & 2002.4 \\
     CBI-M2h & [14] &  304.0 &    0.0 &  400.0 & $ 27.0_{-16.0}^{+ 9.0}$ &  5.0 & 2002.4 \\
     CBI-M2h & [14] &  496.0 &  400.0 &  600.0 & $ 48.0_{- 8.0}^{+ 7.0}$ &  5.0 & 2002.4 \\
     CBI-M2h & [14] &  696.0 &  600.0 &  800.0 & $ 41.0_{- 7.0}^{+ 6.0}$ &  5.0 & 2002.4 \\
     CBI-M2h & [14] &  896.0 &  800.0 & 1000.0 & $ 48.0_{- 7.0}^{+ 6.0}$ &  5.0 & 2002.4 \\
     CBI-M2h & [14] & 1100.0 & 1000.0 & 1200.0 & $ 31.0_{- 8.0}^{+ 6.0}$ &  5.0 & 2002.4 \\
     CBI-M2h & [14] & 1300.0 & 1200.0 & 1400.0 & $ 15.0_{-15.0}^{+10.0}$ &  5.0 & 2002.4 \\
     CBI-M2h & [14] & 1502.0 & 1400.0 & 1600.0 & $ 13.0_{-13.0}^{+12.0}$ &  5.0 & 2002.4 \\
     CBI-M2h & [14] & 1702.0 & 1600.0 & 1800.0 & $  0.0_{- 0.0}^{+22.0}$ &  5.0 & 2002.4 \\
     CBI-M2h & [14] & 1899.0 & 1800.0 & 2000.0 & $  0.0_{- 0.0}^{+17.0}$ &  5.0 & 2002.4 \\
    CBI-M14h & [14] &  304.0 &    0.0 &  400.0 & $ 65.0_{-15.0}^{+12.0}$ &  5.0 & 2002.4 \\
    CBI-M14h & [14] &  496.0 &  400.0 &  600.0 & $ 50.0_{- 9.0}^{+ 7.0}$ &  5.0 & 2002.4 \\
    CBI-M14h & [14] &  696.0 &  600.0 &  800.0 & $ 51.0_{- 7.0}^{+ 7.0}$ &  5.0 & 2002.4 \\
    CBI-M14h & [14] &  896.0 &  800.0 & 1000.0 & $ 41.0_{- 7.0}^{+ 6.0}$ &  5.0 & 2002.4 \\
    CBI-M14h & [14] & 1100.0 & 1000.0 & 1200.0 & $ 27.0_{- 8.0}^{+ 6.0}$ &  5.0 & 2002.4 \\
    CBI-M14h & [14] & 1300.0 & 1200.0 & 1400.0 & $ 30.0_{- 8.0}^{+ 6.0}$ &  5.0 & 2002.4 \\
    CBI-M14h & [14] & 1502.0 & 1400.0 & 1600.0 & $ 34.0_{- 9.0}^{+ 7.0}$ &  5.0 & 2002.4 \\
    CBI-M14h & [14] & 1702.0 & 1600.0 & 1800.0 & $  0.0_{- 0.0}^{+16.0}$ &  5.0 & 2002.4 \\
    CBI-M20h & [14] &  304.0 &    0.0 &  400.0 & $ 60.0_{-16.0}^{+13.0}$ &  5.0 & 2002.4 \\
    CBI-M20h & [14] &  496.0 &  400.0 &  600.0 & $ 49.0_{- 5.0}^{+ 5.0}$ &  5.0 & 2002.4 \\
    CBI-M20h & [14] &  696.0 &  600.0 &  800.0 & $ 36.0_{- 8.0}^{+ 6.0}$ &  5.0 & 2002.4 \\
    CBI-M20h & [14] &  896.0 &  800.0 & 1000.0 & $ 42.0_{- 8.0}^{+ 7.0}$ &  5.0 & 2002.4 \\
    CBI-M20h & [14] & 1100.0 & 1000.0 & 1200.0 & $ 43.0_{- 8.0}^{+ 7.0}$ &  5.0 & 2002.4 \\
    CBI-M20h & [14] & 1300.0 & 1200.0 & 1400.0 & $ 30.0_{-13.0}^{+ 9.0}$ &  5.0 & 2002.4 \\
    CBI-M20h & [14] & 1502.0 & 1400.0 & 1600.0 & $ 37.0_{-13.0}^{+ 9.0}$ &  5.0 & 2002.4 \\
    CBI-M20h & [14] & 1702.0 & 1600.0 & 1800.0 & $ 27.0_{-27.0}^{+11.0}$ &  5.0 & 2002.4 \\
    CBI-M20h & [14] & 1899.0 & 1800.0 & 2000.0 & $ 11.0_{-11.0}^{+18.0}$ &  5.0 & 2002.4 \\
    COBE-DMR & [15] &    2.1 &    2.0 &    2.5 & $  8.5_{- 8.5}^{+16.0}$ &  0.7 & 1996.0 \\
    COBE-DMR & [15] &    3.1 &    2.5 &    3.7 & $ 28.0_{-10.3}^{+ 7.5}$ &  0.7 & 1996.0 \\
    COBE-DMR & [15] &    4.1 &    3.4 &    4.8 & $ 34.0_{- 7.2}^{+ 6.0}$ &  0.7 & 1996.0 \\
    COBE-DMR & [15] &    5.6 &    4.7 &    6.6 & $ 25.1_{- 6.6}^{+ 5.3}$ &  0.7 & 1996.0 \\
    COBE-DMR & [15] &    8.0 &    6.8 &    9.3 & $ 29.4_{- 4.1}^{+ 3.6}$ &  0.7 & 1996.0 \\
    COBE-DMR & [15] &   10.9 &    9.7 &   12.2 & $ 27.7_{- 4.5}^{+ 3.9}$ &  0.7 & 1996.0 \\
    COBE-DMR & [15] &   14.4 &   12.8 &   15.7 & $ 26.1_{- 5.2}^{+ 4.4}$ &  0.7 & 1996.0 \\
    COBE-DMR & [15] &   19.4 &   16.6 &   22.1 & $ 33.0_{- 5.4}^{+ 4.6}$ &  0.7 & 1996.0 \\
        DASI & [16] &  118.0 &  104.0 &  167.0 & $ 61.4_{- 7.1}^{+ 6.3}$ &  4.0 & 2002.2 \\
        DASI & [16] &  203.0 &  173.0 &  255.0 & $ 72.7_{- 3.9}^{+ 3.7}$ &  4.0 & 2002.2 \\
        DASI & [16] &  289.0 &  261.0 &  342.0 & $ 60.5_{- 2.9}^{+ 2.7}$ &  4.0 & 2002.2 \\
        DASI & [16] &  377.0 &  342.0 &  418.0 & $ 40.6_{- 2.5}^{+ 2.4}$ &  4.0 & 2002.2 \\
        DASI & [16] &  465.0 &  418.0 &  500.0 & $ 43.5_{- 2.6}^{+ 2.5}$ &  4.0 & 2002.2 \\
        DASI & [16] &  553.0 &  506.0 &  594.0 & $ 53.3_{- 2.8}^{+ 2.7}$ &  4.0 & 2002.2 \\
        DASI & [16] &  641.0 &  600.0 &  676.0 & $ 40.9_{- 3.4}^{+ 3.2}$ &  4.0 & 2002.2 \\
        DASI & [16] &  725.0 &  676.0 &  757.0 & $ 44.8_{- 4.1}^{+ 3.7}$ &  4.0 & 2002.2 \\
        DASI & [16] &  837.0 &  763.0 &  864.0 & $ 48.2_{- 4.9}^{+ 4.5}$ &  4.0 & 2002.2 \\
        FIRS & [17] &   11.0 &    2.0 &   28.0 & $ 29.4_{- 7.8}^{+ 7.7}$ & -  &1994.7 \\
         IAB & [18] &  120.0 &   65.0 &  221.0 & $ 94.5_{-41.8}^{+41.8}$ & -  &1993.6 \\
      IACB94 & [19] &   33.0 &   17.0 &   59.0 & $111.9_{-60.1}^{+65.4}$ & 14.0 & 1998.3 \\
      IACB94 & [19] &   53.0 &   34.0 &   79.0 & $ 54.6_{-21.9}^{+27.2}$ & 14.0 & 1998.3 \\
      IACB96 & [20] &   39.0 &   15.0 &   77.0 & $ 34.0_{- 6.0}^{+ 8.0}$ & 10.0 & 2001.1 \\
      IACB96 & [20] &   61.0 &   39.0 &   89.0 & $ 40.0_{- 6.0}^{+ 7.0}$ & 10.0 & 2001.1 \\
      IACB96 & [20] &   81.0 &   61.0 &  108.0 & $ 41.0_{- 8.0}^{+ 8.0}$ & 10.0 & 2001.1 \\
      IACB96 & [20] &   99.0 &   81.0 &  123.0 & $ 50.0_{- 9.0}^{+10.0}$ & 10.0 & 2001.1 \\
      IACB96 & [20] &  116.0 &  102.0 &  139.0 & $ 46.0_{- 9.0}^{+10.0}$ & 10.0 & 2001.1 \\
      IACB96 & [20] &  134.0 &  122.0 &  154.0 & $ 56.0_{-10.0}^{+11.0}$ & 10.0 & 2001.1 \\
       JBIAC & [21] &  109.0 &   90.0 &  128.0 & $ 43.0_{-13.3}^{+12.3}$ & -  &1999.8 \\
     MAXIMA1 & [22] &   77.0 &   36.0 &  110.0 & $ 44.7_{- 6.1}^{+ 7.0}$ &  4.0 & 2001.3 \\
     MAXIMA1 & [22] &  147.0 &  111.0 &  185.0 & $ 54.4_{- 5.4}^{+ 5.9}$ &  4.0 & 2001.3 \\
     MAXIMA1 & [22] &  222.0 &  186.0 &  260.0 & $ 78.1_{- 6.0}^{+ 6.5}$ &  4.0 & 2001.3 \\
     MAXIMA1 & [22] &  294.0 &  261.0 &  335.0 & $ 61.9_{- 4.9}^{+ 5.2}$ &  4.0 & 2001.3 \\
     MAXIMA1 & [22] &  381.0 &  336.0 &  410.0 & $ 47.6_{- 5.2}^{+ 5.6}$ &  4.0 & 2001.3 \\
     MAXIMA1 & [22] &  449.0 &  411.0 &  485.0 & $ 38.3_{- 4.5}^{+ 4.8}$ &  4.0 & 2001.3 \\
     MAXIMA1 & [22] &  523.0 &  486.0 &  560.0 & $ 44.0_{- 4.9}^{+ 5.1}$ &  4.0 & 2001.3 \\
     MAXIMA1 & [22] &  597.0 &  561.0 &  635.0 & $ 42.6_{- 5.5}^{+ 5.6}$ &  4.0 & 2001.3 \\
     MAXIMA1 & [22] &  671.0 &  636.0 &  710.0 & $ 45.8_{- 6.4}^{+ 6.4}$ &  4.0 & 2001.3 \\
     MAXIMA1 & [22] &  746.0 &  711.0 &  785.0 & $ 46.8_{- 7.9}^{+ 7.7}$ &  4.0 & 2001.3 \\
     MAXIMA1 & [22] &  856.0 &  786.0 &  935.0 & $ 55.7_{- 7.1}^{+ 6.8}$ &  4.0 & 2001.3 \\
     MAXIMA1 & [22] & 1004.0 &  936.0 & 1085.0 & $ 32.9_{-32.9}^{+15.1}$ &  4.0 & 2001.3 \\
     MAXIMA1 & [22] & 1147.0 & 1086.0 & 1235.0 & $ 14.9_{-14.9}^{+40.0}$ &  4.0 & 2001.3 \\
        MSAM & [23] &   84.0 &   39.0 &  130.0 & $ 35.0_{-11.0}^{+15.0}$ &  5.0 & 2000.2 \\
        MSAM & [23] &  201.0 &  131.0 &  283.0 & $ 49.0_{- 8.0}^{+10.0}$ &  5.0 & 2000.2 \\
        MSAM & [23] &  407.0 &  284.0 &  453.0 & $ 47.0_{- 6.0}^{+ 7.0}$ &  5.0 & 2000.2 \\
        OVRO & [24] &  589.0 &  361.0 &  756.0 & $ 59.0_{- 8.6}^{+ 6.5}$ & -  &2000.2 \\
     PyI-III & [25] &   87.0 &   49.0 &  105.0 & $ 60.0_{- 5.0}^{+ 9.0}$ & 20.0 & 1997.0 \\
     PyI-III & [25] &  170.0 &  120.0 &  239.0 & $ 66.0_{- 9.0}^{+10.7}$ & 20.0 & 1997.0 \\
         PyV & [26] &   50.0 &   21.0 &   94.0 & $ 30.0_{- 4.0}^{+ 4.0}$ & 15.0 & 2001.9 \\
         PyV & [26] &   74.0 &   35.0 &  130.0 & $ 29.0_{- 5.0}^{+ 5.0}$ & 15.0 & 2001.9 \\
         PyV & [26] &  108.0 &   67.0 &  157.0 & $ 33.0_{- 6.0}^{+ 6.0}$ & 15.0 & 2001.9 \\
         PyV & [26] &  140.0 &   99.0 &  185.0 & $ 35.0_{-10.0}^{+10.0}$ & 15.0 & 2001.9 \\
         PyV & [26] &  172.0 &  132.0 &  215.0 & $ 54.0_{-16.0}^{+15.0}$ & 15.0 & 2001.9 \\
         PyV & [26] &  203.0 &  164.0 &  244.0 & $ 90.0_{-23.0}^{+23.0}$ & 15.0 & 2001.9 \\
         PyV & [26] &  233.0 &  195.0 &  273.0 & $ 67.0_{-55.0}^{+33.0}$ & 15.0 & 2001.9 \\
         PyV & [26] &  264.0 &  227.0 &  303.0 & $  0.0_{- 0.0}^{+72.0}$ & 15.0 & 2001.9 \\
        QMAP & [27] &   80.0 &   39.0 &  121.0 & $ 48.3_{- 7.5}^{+ 6.4}$ &  8.0 & 2002.4 \\
        QMAP & [27] &  111.0 &   47.0 &  175.0 & $ 54.6_{- 5.3}^{+ 5.3}$ &  8.0 & 2002.4 \\
        QMAP & [27] &  126.0 &   72.0 &  180.0 & $ 60.9_{- 7.5}^{+ 6.4}$ &  8.0 & 2002.4 \\
          SK & [27] &   87.0 &   58.0 &  126.0 & $ 50.2_{- 5.2}^{+ 8.3}$ & 10.0 & 2002.4 \\
          SK & [27] &  166.0 &  123.0 &  196.0 & $ 70.5_{- 6.2}^{+ 7.3}$ & 10.0 & 2002.4 \\
          SK & [27] &  237.0 &  196.0 &  266.0 & $ 86.8_{- 8.3}^{+10.4}$ & 10.0 & 2002.4 \\
          SK & [27] &  286.0 &  248.0 &  310.0 & $ 87.9_{-10.4}^{+12.5}$ & 10.0 & 2002.4 \\
          SK & [27] &  349.0 &  308.0 &  393.0 & $ 70.4_{-29.1}^{+19.8}$ & 10.0 & 2002.4 \\
    Tenerife & [28] &   20.0 &   12.0 &   30.0 & $ 30.0_{-15.0}^{+11.0}$ & -  &2000.0 \\
      TOCO97 & [27] &   63.0 &   45.0 &   81.0 & $ 35.1_{- 6.4}^{+10.2}$ & 10.0 & 2002.4 \\
      TOCO97 & [27] &   86.0 &   64.0 &  102.0 & $ 43.0_{- 6.3}^{+ 6.9}$ & 10.0 & 2002.4 \\
      TOCO97 & [27] &  114.0 &   90.0 &  134.0 & $ 67.3_{- 5.8}^{+ 6.3}$ & 10.0 & 2002.4 \\
      TOCO97 & [27] &  158.0 &  135.0 &  180.0 & $ 86.4_{- 7.1}^{+ 7.2}$ & 10.0 & 2002.4 \\
      TOCO97 & [27] &  199.0 &  170.0 &  237.0 & $ 82.9_{- 7.6}^{+ 7.6}$ & 10.0 & 2002.4 \\
      TOCO98 & [27] &  128.0 &   95.0 &  154.0 & $ 53.7_{-16.3}^{+18.1}$ &  8.0 & 2002.4 \\
      TOCO98 & [27] &  152.0 &  114.0 &  178.0 & $ 80.6_{-10.8}^{+10.8}$ &  8.0 & 2002.4 \\
      TOCO98 & [27] &  226.0 &  170.0 &  263.0 & $ 81.6_{- 7.9}^{+ 6.9}$ &  8.0 & 2002.4 \\
      TOCO98 & [27] &  306.0 &  247.0 &  350.0 & $ 68.8_{-10.8}^{+ 9.8}$ &  8.0 & 2002.4 \\
      TOCO98 & [27] &  409.0 &  344.0 &  451.0 & $ 23.3_{-22.4}^{+22.4}$ &  8.0 & 2002.4 \\
       VIPER & [29] &  108.0 &   30.0 &  228.0 & $ 61.0_{-22.0}^{+31.0}$ &  8.0 & 2000.3 \\
       VIPER & [29] &  173.0 &   73.0 &  288.0 & $ 77.0_{-20.0}^{+26.0}$ &  8.0 & 2000.3 \\
       VIPER & [29] &  237.0 &  126.0 &  336.0 & $ 65.0_{-17.0}^{+24.0}$ &  8.0 & 2000.3 \\
       VIPER & [29] &  263.0 &  150.0 &  448.0 & $ 79.0_{-14.0}^{+18.0}$ &  8.0 & 2000.3 \\
       VIPER & [29] &  422.0 &  291.0 &  604.0 & $ 28.0_{-15.0}^{+15.0}$ &  8.0 & 2000.3 \\
       VIPER & [29] &  589.0 &  448.0 &  796.0 & $ 65.0_{-25.0}^{+25.0}$ &  8.0 & 2000.3 \\
         VSA & [30] &  160.0 &  100.0 &  190.0 & $ 62.2_{-10.0}^{+11.7}$ &  3.5 & 2002.9 \\
         VSA & [30] &  220.0 &  190.0 &  250.0 & $ 76.8_{- 9.3}^{+10.0}$ &  3.5 & 2002.9 \\
         VSA & [30] &  289.0 &  250.0 &  310.0 & $ 73.4_{- 7.8}^{+ 8.3}$ &  3.5 & 2002.9 \\
         VSA & [30] &  349.0 &  310.0 &  370.0 & $ 51.0_{- 5.7}^{+ 5.5}$ &  3.5 & 2002.9 \\
         VSA & [30] &  416.0 &  370.0 &  450.0 & $ 41.8_{- 4.4}^{+ 4.0}$ &  3.5 & 2002.9 \\
         VSA & [30] &  479.0 &  450.0 &  500.0 & $ 40.5_{- 7.4}^{+ 7.3}$ &  3.5 & 2002.9 \\
         VSA & [30] &  537.0 &  500.0 &  580.0 & $ 53.5_{- 5.4}^{+ 5.3}$ &  3.5 & 2002.9 \\
         VSA & [30] &  605.0 &  580.0 &  640.0 & $ 38.2_{- 8.0}^{+ 7.7}$ &  3.5 & 2002.9 \\
         VSA & [30] &  670.0 &  640.0 &  700.0 & $ 47.3_{- 6.8}^{+ 6.8}$ &  3.5 & 2002.9 \\
         VSA & [30] &  726.0 &  700.0 &  750.0 & $ 43.8_{- 9.5}^{+ 8.3}$ &  3.5 & 2002.9 \\
         VSA & [30] &  795.0 &  750.0 &  850.0 & $ 59.9_{- 5.6}^{+ 5.2}$ &  3.5 & 2002.9 \\
         VSA & [30] &  888.0 &  850.0 &  950.0 & $ 38.4_{- 7.9}^{+ 7.6}$ &  3.5 & 2002.9 \\
         VSA & [30] & 1002.0 &  950.0 & 1050.0 & $  0.0_{- 0.0}^{+33.0}$ &  3.5 & 2002.9 \\
         VSA & [30] & 1119.0 & 1050.0 & 1200.0 & $ 33.5_{-11.6}^{+ 9.1}$ &  3.5 & 2002.9 \\
         VSA & [30] & 1271.0 & 1200.0 & 1350.0 & $  0.0_{- 0.0}^{+37.8}$ &  3.5 & 2002.9 \\
         VSA & [30] & 1419.0 & 1350.0 & 1700.0 & $ 36.2_{-31.5}^{+17.2}$ &  3.5 & 2002.9 \\
\enddata 
\tablerefs{[1] \cite{kuo02}, [2] \cite{tanaka96}, \cite{lineweaver98}, [3] \cite{benoit02}, [4] \cite{gunderson95}, [5] \cite{debernardis94}, [6] \cite{masi96}, [7] \cite{tucker97}, [8] \cite{mauskopf00}, [9] \cite{ruhl02}, [10] \cite{scott96}, [11] \cite{baker99}, [12] \cite{padin01}, [13] \cite{mason02}, [14] \cite{pearson02}, [15] \cite{tegmark97}, [16] \cite{halverson02}, [17] \cite{ganga94a}, [18] \cite{piccirillo93}, [19] \cite{femenia98}, [20] \cite{romeo01}, [21] \cite{dicker99}, [22] \cite{lee01}, [23] \cite{wilson00}, [24] \cite{leitch00}, [25] \cite{platt97}, [26] \cite{coble01}, [27] \cite{miller02}, [28] \cite{gutierrez00}, [29] \cite{peterson00}, [30] \cite{grainge02}.}
\tablenotetext{a}{The 1$\sigma$ calibration uncertainty in temperature, $\sigma_u$, is given as a percentage and allows the data points taken at the same time using the same instrument to shift upwards or downwards together.  For observations that result in a single data point, $\sigma_u$ is not given since either it is not quoted in the literature, or it has been treated by adding it in quadrature to the statistical error bars.}
\end{deluxetable}
\clearpage
\begin{deluxetable}{lccccccccccccc}
\rotate
\tabletypesize{\tiny}
\tablecolumns{8} 
\tablewidth{0pc}
\setlength{\tabcolsep}{0.04in}
\tablecaption{\label{obstech} Details of the CMB observational techniques.} 
\tablehead{ 
\colhead{Experiment} & \colhead{Ref.}   & \colhead{$\nu_{\rm main}$}  & \colhead{$\Delta \nu_{\rm main}$} & \colhead{$\nu$ Range}  & \colhead{$\theta_{\rm beam}$ FWHM}  & \colhead{Point. Uncert.$^a$} & 
\colhead{Area of} & \colhead{Calibration}  & \colhead{Instrument}  & \colhead{Platform}  & \colhead{Inst. Alt.}   & \colhead{$|$Gal. Lat.$|$}    & \colhead{Gal. Long.}\\
\colhead{} & \colhead{}   & \colhead{(GHz)}  & \colhead{(GHz)} & \colhead{(GHz)}  & \colhead{(arcmin)}  & \colhead{(arcmin)} & 
\colhead{Sky (deg$^2$)} & \colhead{Source}  & \colhead{Type}  & \colhead{}  & \colhead{(m)}   & \colhead{Range (deg)}    & \colhead{Range (deg)}}
\startdata 
       ACBAR &  [1] & 150.0 & 15.0 & 150.0 - 280.0 &   4.5 &  0.30 &    24.8 & Venus \& Mars &  Bolometer &     Ground &   2.9 &  36.7 -  57.0 & 250.3 - 276.5 \\
    ACME-MAX &  [2] & 180.0 &  7.0 & 105.0 - 420.0 &  30.0 &  1.00 &    18.8 &       Jupiter &  Bolometer &    Balloon &  35.4 &  40.8 -  76.6 &  67.4 - 108.5 \\
   ACME-SP91 &  [3] &  27.7 &  1.2 &  27.7 -  27.7 &  96.0 &  5.00 &    44.0 &      Taurus A &       HEMT &     Ground &   2.8 &  45.0 -  55.0 & 275.0 - 305.0 \\
   ACME-SP94 &  [4] &  35.0 &  1.2 &  27.7 -  41.5 &  83.0 &  7.20 &    20.0 &      H/C Load &       HEMT &     Ground &   2.8 &  40.0 -  55.0 & 270.0 - 290.0 \\
    ARCHEOPS &  [5] & 190.0 & 31.0 & 143.0 - 545.0 &   8.0 &  1.50 &  5198.0 &        Dipole &  Bolometer &    Balloon &  34.0 &  30.0 -  90.0 &  30.0 - 220.0 \\
       ARGO1 &  [6] & 150.0 & 15.0 & 150.0 - 600.0 &  52.0 &  2.00 &    26.0 &      H/C Load &  Bolometer &    Balloon &  40.0 &  22.0 -  35.0 &  64.0 -  79.0 \\
       ARGO2 &  [7] & 150.0 & 15.0 & 150.0 - 600.0 &  52.0 &  2.00 &    10.0 &      H/C Load &  Bolometer &    Balloon &  40.0 &   0.0 -   7.8 &  33.8 -  35.7 \\
         BAM &  [8] & 147.0 & 20.0 & 111.0 - 255.0 &  42.0 &  3.00 &     1.0 &       Jupiter &    Bol/Int &    Balloon &  41.5 &  12.8 -  32.8 & 100.3 - 108.0 \\
      BOOM97 &  [9] & 153.0 & 21.0 &  96.0 - 153.0 &  18.0 &  1.00 &   365.0 &       Jupiter &  Bolometer &    Balloon &  38.5 &  11.0 -  83.0 &  17.0 - 178.0 \\
      BOOM98 & [10] & 150.0 & 11.0 &  90.0 - 410.0 &  11.1 &  2.50 &  1213.0 &        Dipole &  Bolometer &    Balloon &  39.0 &  18.0 -  45.0 & 239.0 - 266.0 \\
        CAT1 & [11] &  16.5 &  0.2 &  13.5 -  16.5 & 117.6 & - &     4.0 &         Cas A &   HEMT/Int &     Ground &   0.0 &  29.5 -  38.0 & 140.5 - 151.9 \\
        CAT2 & [12] &  16.5 &  0.2 &  13.5 -  16.5 & 117.6 & - &     4.0 &         Cas A &   HEMT/Int &     Ground &   0.0 &  33.9 -  40.4 &  90.1 -  98.9 \\
      CBI-1a & [13] &  31.0 &  0.5 &  26.5 -  35.5 &   3.0 &  0.03 &     1.1 &      Taurus A &   HEMT/Int &     Ground &   5.1 &  23.3 -  25.0 & 229.7 - 230.9 \\
      CBI-1b & [13] &  31.0 &  0.5 &  26.5 -  35.5 &   3.0 &  0.03 &     1.1 &      Taurus A &   HEMT/Int &     Ground &   5.1 &  47.8 -  49.1 & 348.0 - 350.3 \\
     CBI-D8h & [14] &  31.0 &  0.5 &  26.5 -  35.5 &   3.0 &  0.03 &     1.1 &       Jupiter &   HEMT/Int &     Ground &   5.1 &  23.3 -  25.0 & 229.7 - 230.9 \\
    CBI-D14h & [14] &  31.0 &  0.5 &  26.5 -  35.5 &   3.0 &  0.03 &     1.1 &       Jupiter &   HEMT/Int &     Ground &   5.1 &  53.4 -  54.8 & 356.4 - 358.7 \\
    CBI-D20h & [14] &  31.0 &  0.5 &  26.5 -  35.5 &   3.0 &  0.03 &     1.1 &       Jupiter &   HEMT/Int &     Ground &   5.1 &  27.6 -  29.3 &  43.9 -  45.0 \\
     CBI-M2h & [15] &  31.0 &  0.5 &  26.5 -  35.5 &   3.0 &  0.05 &    40.0 &       Jupiter &   HEMT/Int &     Ground &   5.1 &  53.0 -  55.0 & 175.0 - 177.0 \\
    CBI-M14h & [15] &  31.0 &  0.5 &  26.5 -  35.5 &   3.0 &  0.05 &    40.0 &       Jupiter &   HEMT/Int &     Ground &   5.1 &  48.0 -  50.0 & 347.0 - 351.0 \\
    CBI-M20h & [15] &  31.0 &  0.5 &  26.5 -  35.5 &   3.0 &  0.05 &    40.0 &       Jupiter &   HEMT/Int &     Ground &   5.1 &  26.0 -  28.0 &  42.0 -  45.0 \\
    COBE-DMR & [16] &  53.0 &  0.1 &  31.5 -  90.0 & 420.0 &  3.00 & 41253.0 &      H/C Load &       HEMT &  Satellite & 900.0 &  20.0 -  90.0 &   0.0 - 360.0 \\
        DASI & [17] &  31.0 &  0.5 &  26.5 -  35.5 &  20.0 &  2.00 &   400.0 &  Gal. Sources &   HEMT/Int &     Ground &   2.8 &  26.9 -  67.3 & 254.7 - 324.4 \\
        FIRS & [18] & 167.0 & 19.0 & 167.0 - 682.0 & 228.0 & 60.00 & 14000.0 &        Dipole &  Bolometer &    Balloon &  36.3 &   0.0 -  80.0 &  45.0 - 200.0 \\
         IAB & [19] & 136.0 &  1.5 & 136.0 - 136.0 &  50.0 &  2.00 &     6.0 &      H/C Load &  Bolometer &     Ground &   3.3 &  22.2 -  32.1 & 297.4 - 308.4 \\
      IACB94 & [20] & 116.0 &  1.5 &  90.9 - 272.7 & 121.8 &  5.40 &   150.0 &          Moon &  Bolometer &     Ground &   2.4 &   0.0 -  49.2 &  63.5 - 128.9 \\
      IACB96 & [21] & 142.9 &  1.5 &  96.8 - 272.7 &  81.0 &  5.40 &  1000.0 &          Moon &  Bolometer &     Ground &   2.4 &   0.0 -  87.8 &  41.3 - 204.1 \\
       JBIAC & [22] &  33.0 &  1.5 &  33.0 -  33.0 & 120.0 & - &   260.0 &          Moon &   HEMT/Int &     Ground &   2.4 &   0.0 -  76.0 &  65.0 - 181.0 \\
     MAXIMA1 & [23] & 150.0 & 35.0 & 150.0 - 410.0 &  10.0 &  0.95 &    57.0 &        Dipole &  Bolometer &    Balloon &  37.0 &  44.0 -  54.3 &  84.7 - 100.5 \\
        MSAM & [24] & 170.0 & 22.5 & 170.0 - 680.0 &  30.0 &  2.50 &    10.0 &       Jupiter &  Bolometer &    Balloon &  39.0 &  25.0 -  36.0 & 113.0 - 120.0 \\
        OVRO & [25] &  31.7 &  3.0 &  14.5 -  31.7 &   7.4 &  2.00 &     6.0 &       Jupiter &       HEMT &     Ground &   1.2 &  25.1 -  29.3 & 120.6 - 125.3 \\
     PyI-III & [26] &  90.0 & 18.0 &  90.0 -  90.0 &  45.0 &  6.00 &   121.0 &      H/C Load &  Bolometer &     Ground &   2.8 &  60.0 -  70.0 & 270.0 - 355.0 \\
         PyV & [27] &  40.3 &  2.8 &  40.3 -  40.3 &  60.0 &  9.00 &   598.0 &      H/C Load &       HEMT &     Ground &   2.8 &  30.0 -  70.0 & 270.0 - 355.0 \\
        QMAP & [28] &  37.0 &  3.1 &  31.0 -  42.0 &  48.0 &  3.60 &   527.0 &         Cas A &       HEMT &    Balloon &  30.0 &   8.0 -  46.0 & 102.0 - 132.0 \\
          SK & [29] &  42.0 &  3.5 &  31.0 -  42.0 &  28.0 &  1.80 &   200.0 &         Cas A &       HEMT &     Ground &   0.5 &  19.0 -  35.0 & 114.0 - 132.0 \\
    Tenerife & [30] &  15.0 &  0.8 &  10.0 -  15.0 & 300.0 & - &  2000.0 &   Combination &       HEMT &     Ground &   2.4 &  40.0 -  90.0 &  40.0 - 210.0 \\
      TOCO97 & [31] &  37.0 &  3.0 &  31.0 - 144.0 &  48.0 &  0.45 &   600.0 &       Jupiter &   HEMT/SIS &     Ground &   5.2 &   0.0 -  55.0 & 270.0 - 335.0 \\
      TOCO98 & [31] & 144.0 &  1.7 &  31.0 - 144.0 &  12.0 &  0.45 &   600.0 &       Jupiter &   HEMT/SIS &     Ground &   5.2 &   0.0 -  55.0 & 270.0 - 335.0 \\
       VIPER & [32] &  40.0 &  3.0 &  40.0 -  40.0 &  15.6 &  4.00 &     4.0 &      H/C Load &       HEMT &     Ground &   2.8 &  50.0 -  60.0 & 330.0 - 342.0 \\
         VSA & [33] &  34.0 &  0.8 &  34.0 -  34.0 & 120.0 &  5.00 &   140.0 &       Jupiter &   HEMT/Int &     Ground &   2.4 &  31.5 -  54.3 & -$^b$ \\
\enddata 
\tablerefs{[1] \cite{kuo02}, [2] \cite{alsop92}, \cite{lim96}, \cite{tanaka96}, [3] \cite{gunderson95}, [4] \cite{benoit02} [5] \cite{ganga94b}, \cite{gunderson95}, [6] \cite{debernardis93}, \cite{debernardis94}, [7] \cite{debernardis93}, \cite{debernardis94}, \cite{masi95}, \cite{masi96}, [8] \cite{tucker97}, [9] \cite{mauskopf00}, \cite{piacentini02}, [10] \cite{crill02}, \cite{netterfield02}, \cite{ruhl02}, [11] \cite{scott96}, [12] \cite{baker99}, [13] \cite{padin01}, \cite{padin02}, [14] \cite{mason02}, [15] \cite{mason02}, \cite{pearson02}, [16] \cite{kogut92}, \cite{kogut96}, \cite{tegmark97}, [17] \cite{halverson02}, \cite{leitch02}, [18] \cite{page90}, \cite{meyer91}, \cite{ganga94a}, [19] \cite{piccirillo93}, [20] \cite{femenia98}, [21] \cite{femenia98}, \cite{romeo01}, [22] \cite{dicker99}, \cite{melhuish99}, [23] \cite{lee01}, \cite{hanany00}, [24] \cite{fixsen96}, \cite{wilson00}, [25] \cite{leitch00}, [26] \cite{dragovan94}, \cite{ruhl95}, \cite{platt97}, [27] \cite{coble99}, \cite{coble01}, [28] \cite{deoliveira98}, \cite{devlin98}, \cite{herbig98}, \cite{miller02}, [29] \cite{netterfield97}, \cite{miller02}, [30] \cite{davies96}, \cite{gutierrez00}, [31] \cite{miller02}, [32] \cite{peterson00}, [33] \cite{grainge02}, \cite{scott02}, \cite{taylor02}, \cite{watson02}.}
\tablenotetext{a}{Where pointing uncertainties are not given, they are not quoted in the literature.}
\tablenotetext{b}{The VSA results quoted in the literature are the combined detections from a number of separate fields observed at various galactic longitudes and latitudes.  Information is not given in the literature to enable the contributions from the different fields to be separated. Therefore, the VSA galactic longitude range is omitted.  However, since the fields are not very dispersed in galactic latitude, this range is listed.}
\end{deluxetable}
\clearpage
\begin{deluxetable}{lccccccccccccccc} 
\rotate
\tabletypesize{\tiny}
\tablecolumns{16} 
\tablewidth{0pc} 
\setlength{\tabcolsep}{0.04in}
\tablecaption{\label{resres} Results of the residual analyses.}
\tablehead{ 
\colhead{} & \colhead{$\ell_{\rm eff}$} & \colhead{$\begin{array}{c}|b| \\ {\rm range}^a\end{array}$} & \colhead{$\begin{array}{c}\rm central \\|b|^b \end{array}$} & \colhead{$\nu_{\rm main}$} & \colhead{$\frac{(\nu_{\rm max}-\nu_{\rm min})}{\nu_{\rm main}}$} & \colhead{$\begin{array}{c}\rm cal.\\ \rm source\end{array}$} & \colhead{$\begin{array}{c}\rm inst.\\ \rm type\end{array}$} & \colhead{$\frac{\theta_{\rm beam}}{\ell_{\rm eff}}$} & \colhead{$\begin{array}{c}\rm point.\\ \rm uncert\end{array}$} & \colhead{$\begin{array}{c}\rm area \\ \rm of sky\end{array}$} &\colhead{$\begin{array}{c}\rm inst. \\ \rm platform\end{array}$}& \colhead{$\begin{array}{c}\rm inst.\\ \rm alt.\end{array}$} & \colhead{$\frac{\Delta\ell}{ell_{eff}}$} & \colhead{$\begin{array}{c}\rm gal.\\ \rm long.\end{array}$}& \colhead{$\frac{\Delta\nu_{\rm main}}{\nu_{\rm main}}$}}
\startdata
$\chi^2$ per dof & 0.98 & 0.88 & 1.00 & 1.00 & 1.00 & 1.00 & 1.00 &1.00&1.01& 0.99 & 1.00 &0.99&1.00&0.98 & 1.00 \\
significance$^c$ & $>1\sigma$ & $>2\sigma$ &$<\frac{1}{2}\sigma$ &$<\frac{1}{2}\sigma$ &$<\frac{1}{2}\sigma$ &$ <\frac{1}{2}\sigma$&$<\frac{1}{2}\sigma$&$<\frac{1}{2}\sigma$&$<\frac{1}{2}\sigma$&$<\frac{1}{2}\sigma$&$<\frac{1}{2}\sigma$& $<\frac{1}{2}\sigma$ & $ <\frac{1}{2}\sigma$& $<\frac{1}{2}\sigma$& $<\frac{1}{2}\sigma$\\
\enddata 
\tablenotetext{a}{The $|b|$ range analysis involves a 2 dimensional fit (see \S5).}
\tablenotetext{b}{The central $|b|$ analysis involves a 1 dimensional fit (see \S5).}
\tablenotetext{c}{The significance of the deviation of the zero-line from the best-fit linear model.} 
\end{deluxetable}

\end{document}